\documentclass[twocolumn]{aastex63}

\submitjournal{ApJ}

\shorttitle{M17 magnetic field}
\shortauthors{Thuong Hoang et al.}

\usepackage{chngcntr}
\usepackage{graphicx}
\graphicspath{{./}{figures/}}
\usepackage{multirow}
\usepackage{comment}
\usepackage{natbib}
\usepackage{amsmath,amsbsy}
\newcommand*\Bell{\ensuremath{\boldsymbol\ell}}

\begin{document}

\title{Studying magnetic fields and dust in M17 using polarized thermal dust emission observed by SOFIA/HAWC+} %\footnote{January, 27th, 2021} }

\correspondingauthor{Thuong Duc Hoang}
\email{hoang-duc.thuong@usth.edu.vn}

\author[0000-0002-3437-5228]{Thuong Duc Hoang}
\affiliation{University of Science and Technology of Hanoi (USTH), Vietnam Academy of Science and Technology (VAST), 18 Hoang Quoc Viet, Hanoi, Vietnam}

\author[0000-0002-5913-5554]{Nguyen Bich Ngoc}
\affiliation{Department of Astrophysics, Vietnam National Space Center, Vietnam Academy of Science and Technology, 18 Hoang Quoc Viet, Hanoi, Vietnam}
\affiliation{Graduate University of Science and Technology, Vietnam Academy of Science and Technology, 18 Hoang Quoc Viet, Hanoi, Vietnam}

\author[0000-0002-2808-0888]{Pham Ngoc Diep}
\affiliation{Department of Astrophysics, Vietnam National Space Center, Vietnam Academy of Science and Technology, 18 Hoang Quoc Viet, Hanoi, Vietnam}

\author[0000-0002-6488-8227]{Le Ngoc Tram}
\affiliation{Stratospheric Observatory for Infrared Astronomy, Universities Space Research Association, NASA Ames Research Center, MS 232-11, Moffett Field, 94035 CA, USA}
\affiliation{Max-Planck-Institut für Radioastronomie, Auf dem Hügel 69, 53-121, Bonn, Germany}

\author[0000-0003-2017-0982]{Thiem Hoang}
\affiliation{Korea Astronomy and Space Science Institute, 776 Daedeokdae-ro, Yuseong-gu, Daejeon 34055, Republic of Korea}
\affiliation{University of Science and Technology, Korea, 217 Gajeong-ro, Yuseong-gu, Daejeon 34113, Republic of Korea}

\author[0000-0003-4243-6809]{Wanggi Lim}
\affiliation{Stratospheric Observatory for Infrared Astronomy, Universities Space Research Association, NASA Ames Research Center, MS 232-11, Moffett Field, 94035 CA, USA}

\author[0000-0003-1990-1717]{Ngan Le}\affiliation{Institute of Astronomy, Faculty of Physics, Astronomy and Informatics, Nicolaus Copernicus University, Grudziadzka 5, 87-100 Torun, Poland}

\author[0000-0002-8557-3582]{Kate Pattle}
\affiliation{National University of Ireland Galway, University Road, Galway, Ireland H91 TK33}
\affiliation{Department of Physics and Astronomy, University College London, Gower Street, London WC1E 6BT, United Kingdom}

\author[0000-0002-5678-1008]{Dieu D. Nguyen}
\affiliation{Department of Physics, International University, Quarter 6, Linh Trung Ward, Thu Duc City, Ho Chi Minh City, Vietnam}
\affiliation{Vietnam National University, Ho Chi Minh City, Vietnam}

\author{Nguyen Thi Phuong}
\affiliation{Korea Astronomy and Space Science Institute, 776 Daedeokdae-ro, Yuseong-gu, Daejeon 34055, Republic of Korea}
\affiliation{Department of Astrophysics, Vietnam National Space Center, Vietnam Academy of Science and Technology, 18 Hoang Quoc Viet, Hanoi, Vietnam}

\author[0000-0002-6372-8395]{Nguyen Fuda}
\affiliation{Department of Physics, International University, Quarter 6, Linh Trung Ward, Thu Duc City, Ho Chi Minh City, Vietnam}
\affiliation{Vietnam National University, Ho Chi Minh City, Vietnam}

\author{Tuan Van Bui}
\affiliation{University of Science and Technology of Hanoi (USTH), Vietnam Academy of Science and Technology (VAST), 18 Hoang Quoc Viet, Hanoi, Vietnam}

\author[0000-0001-9654-8051]{Gia Bao Truong Le}
\affiliation{Department of Physics, International University, Quarter 6, Linh Trung Ward, Thu Duc City, Ho Chi Minh City, Vietnam}
\affiliation{Vietnam National University, Ho Chi Minh City, Vietnam}

\author[0000-0003-4935-462X]{Hien Phan}
\affiliation{University of Science and Technology of Hanoi (USTH), Vietnam Academy of Science and Technology (VAST), 18 Hoang Quoc Viet, Hanoi, Vietnam}

\author{Nguyen Chau Giang}
\affiliation{Korea Astronomy and Space Science Institute, 776 Daedeokdae-ro, Yuseong-gu, Daejeon 34055, Republic of Korea}

%%%%%%%%%%%%%%%%%%%%%%%%%%%%%%%%%%%
\begin{abstract}
We report the highest spatial resolution measurement of magnetic fields (B-field) in M17 using thermal dust polarization taken by SOFIA/HAWC+ centered at 154 $\mu$m wavelength. Using the Davis-Chandrasekhar-Fermi method, we found the presence of strong B-fields of $980 \pm 230\;\mu$G and $1665 \pm 885\;\mu$G in lower-density (M17-N) and higher-density (M17-S) regions, respectively. The B-field morphology in M17-N possibly mimics the fields in gravitational collapse molecular cores while in M17-S the fields run perpendicular to the density structure and display a pillar and an asymmetric large-scale hourglass shape. The mean B-field strengths are used to determine the Alfv\'enic Mach numbers, revealing B-fields dominate turbulence. We calculate the mass-to-flux ratio, $\lambda$, and obtain $\lambda=0.07$ for M17-N and $0.28$ for M17-S. The sub-critical values of $\lambda$ are in agreement with the lack of massive stars formed in M17. To study dust physics, we analyze the relationship between dust polarization fraction, $p$, and emission intensity, $I$, gas column density, $N({\rm H_2})$, polarization angle dispersion function $S$, and dust temperature, $T_{\rm d}$. $p$ decreases with intensity as $I^{-\alpha}$ with $\alpha = 0.51$. $p$ also decreases with increasing $N(\rm H_{2})$, which can be explained by the decrease of grain alignment by radiative torques (RATs) toward denser regions with a weaker radiation field and/or tangling of magnetic fields. $p$ tends to first increase with $T_{\rm d}$ and then decreases at higher $T_ {\rm d}$. The latter feature seen in M17-N at high $T_{d}$ when $N(\rm H_{2})$ and $S$ decrease is evidence of the RAT disruption effect.
\end{abstract}

%%%%%%%%%%%%%%%%%%%%%%%%%%%%%%%%%%%%%%%%%%%%%
\section{Introduction}\label{sec:intro}
Star formation is a complex process that involves self-gravity, turbulence, magnetic fields, and stellar feedback. Understanding the exact role of magnetic fields in the evolution of molecular clouds (MCs) and star formation process is a challenge of modern astrophysics. In the past decades, there are emerging evidences suggesting the importance of magnetic fields (B-fields) in the evolution of MCs and star formation \citep{McKee, book_role_mag}. There are two types of B-field models in which B-fields play contrasting roles. First, the strong B-field models support a paradigm of magnetic pressure acting against gravitational collapse of MCs. This magnetic-driven support is suppressed when the ratio of the core mass to the magnetic flux exceeds a critical value that turns the cloud into a state of gravitational collapse to form a new star \citep{Nakano_1978}. Second, in the weak-field models, B-fields are sufficiently weak in MCs and dominated by turbulence. Star formation takes place in filaments that are likely produced by intersection of turbulent supersonic flows \citep[e.g.,][]{Padoan_1999, Elmegreen_2000, MacLow2004, crutcher_richard}. It is, thus, necessary to carry out observations of B-fields in specific MCs (e.g., M17 in this work) to investigate their effects on star formation and the subsequent evolution of the entire clouds to testify theoretical predictions \citep{Seifried_2015, Federrath_2016, Li_P_S_2019, book_role_mag}.

Dust polarization induced by aligned grains is widely used to map B-fields \citep[see e.g.,][]{crutcher_richard}. This method is based on the fact that the polarization direction of thermal dust emission is perpendicular to the B-fields. The strength of B-fields can then be estimated using the Davis-Chandrasekhar-Fermi (DCF) method \citep{Davis1951, C_F_1953}. The DCF method measures indirectly the strength based on the fluctuations of B-fields that are encoded in the dispersion of dust polarization directions. Although we have a full-sky map of dust polarization provided by {\it Planck} at 353~GHz \citep[see e.g.,][]{planck2015_dust_pol} and many studies of B-fields at MC scales \citep[see e.g., filament structures and star-forming cores;][]{Dotson1996, Houde_2002, Pellegrini_2007, Chen_2012, Pattle_2017, Pattle_2018, Wurster_2018, Chuss_2019, Hennebelle_2019, Sugitani_2019}, dust polarization data at higher resolutions are still lacking. Recently, we start to have high-resolution polarisation observations made with the Atacama Large Millimeter/sub-Millimeter Array \citep{Liu_2020, Beuther_2020, Sanhueza_2021}.

Dust polarization also allows us to get insights into fundamental properties of dust grains such as grain shapes, size distribution, and alignment. Grain alignment is a longstanding problem of astrophysics, and the leading theory for grain alignment is the Radiative Torque Alignment theory (RATA; \citealt{Draine_1997}; \citealt{Lazarian2007}; \citealt{hoanglazarian2016}) \citep[see reviews e.g.,][]{, Andersson_2015, lazarian2016grain}. Note that radiative torques (RATs) arising from the interaction of an anisotropic radiation field with an irregular grain were first suggested by \cite{Dolginov1976} and later numerically demonstrated by \cite{Draine_1996}, and analytically modelled by \cite{Lazarian2007}. The grain alignment efficiency by RATs is found to increase with increasing radiation field or dust temperature (see \citealt{Hoang_2021}), which results in the increase of the polarization fraction with the dust temperature \citep{Lee_2020}. Furthermore, \cite{Hoang_2019} realized that large grains will be disrupted and depleted around a very strong radiation source via a mechanism, the so-called Radiative Torque Disruption (RATD; see \citealt{Hoang2020} for a review). The basic idea of the RATD mechanism is that an intense radiation field can spin up dust grains to extremely fast rotation, such that the centrifugal stress can exceed the tensile strength of the grain material and disrupts the dust grain into smaller fragments. The RATD effect is found to decrease the polarization fraction predicted by the RATA theory \citep{Lee_2020}. The combination of RATA and RATD was demonstrated to successfully reproduce the observed dust polarization data in various regions of strong radiation fields such as Oph-A \citep{Tram2021c}, 30 Doradus \citep{tram2021sofia}. Therefore, dust polarization observations toward strong radiation sources like M17 are crucial to test grain alignment and disruption by RATs.
\begin{figure*}[htbp!]
\includegraphics[width=0.96\textwidth]{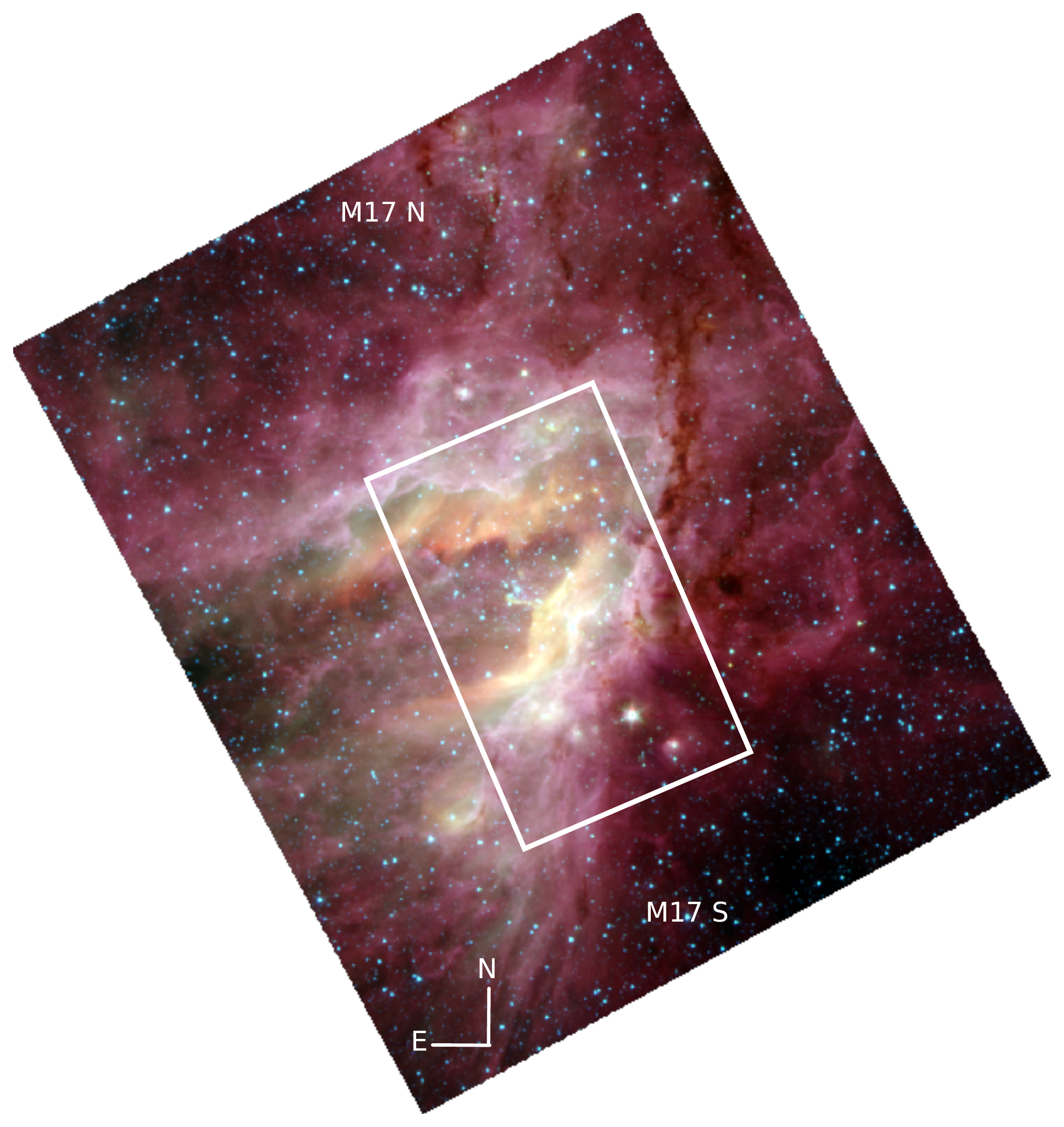}
\caption{An RGB image of M17 observed by {\it Spitzer} (R = 8 $ \rm \mu m$, G = 4.5 $\rm \mu m$, and B = 3.6 $\rm \mu m$). The white rectangle is the observed region of SOFIA/HAWC+ used in this work.}
\label{fig1:M17_RGB}
\end{figure*}

 M17 is a well-known star-forming region \citep{Povich_2009, Lim_2020} which is located in the Omega Nebula or Swan Nebula (also known as Horseshoe Nebula) in the constellation of Sagittarius at a distance of \mbox{1.98 kpc} \citep{Xu_2011}. Figure \ref{fig1:M17_RGB} is a RGB image of the region using {\it Spitzer}\footnote{https://www.spitzer.caltech.edu} Galactic Legacy Infrared Midplane Survey Extraordinaire (GLIMPSE) mosaic data. Besides the study of B-fields and dust physics, since M17 is the closest giant \ion{H}{2} region to Earth it is thus an excellent laboratory for the investigation of stellar feedback from a nearby massive star cluster and a photodissociation region (PDR) as well as infrared sources in the regions such as UC 1, IRS 5, CEN 92, Anon 1, and Anon 3 \citep{Lim_2020}. In the present work, we use data taken by the High-resolution Airborne Wideband Camera Plus \citep[HAWC+;][]{Harper_2018} accommodated on Stratospheric Observatory for Infrared Astronomy \citep[SOFIA;][]{Temi_sofia_2018} to serve our purposes. 

The structure of the paper is organized as follows. The SOFIA/HAWC+ observations of M17 are presented in Section~\ref{sec:obs}. In Section~\ref{sec:analysis}, we describe the use of the DCF method to estimate the strengths of B-fields. We then present the obtained results, including the B-field morphologies and strengths, and discuss the implications of dust polarization fraction for grain alignment and disruption in Section~\ref{subsec:result}. A summary of our main findings is presented in Section~\ref{subsec:summary}.

\begin{figure*}[htbp!]
\includegraphics[width=1.\textwidth]{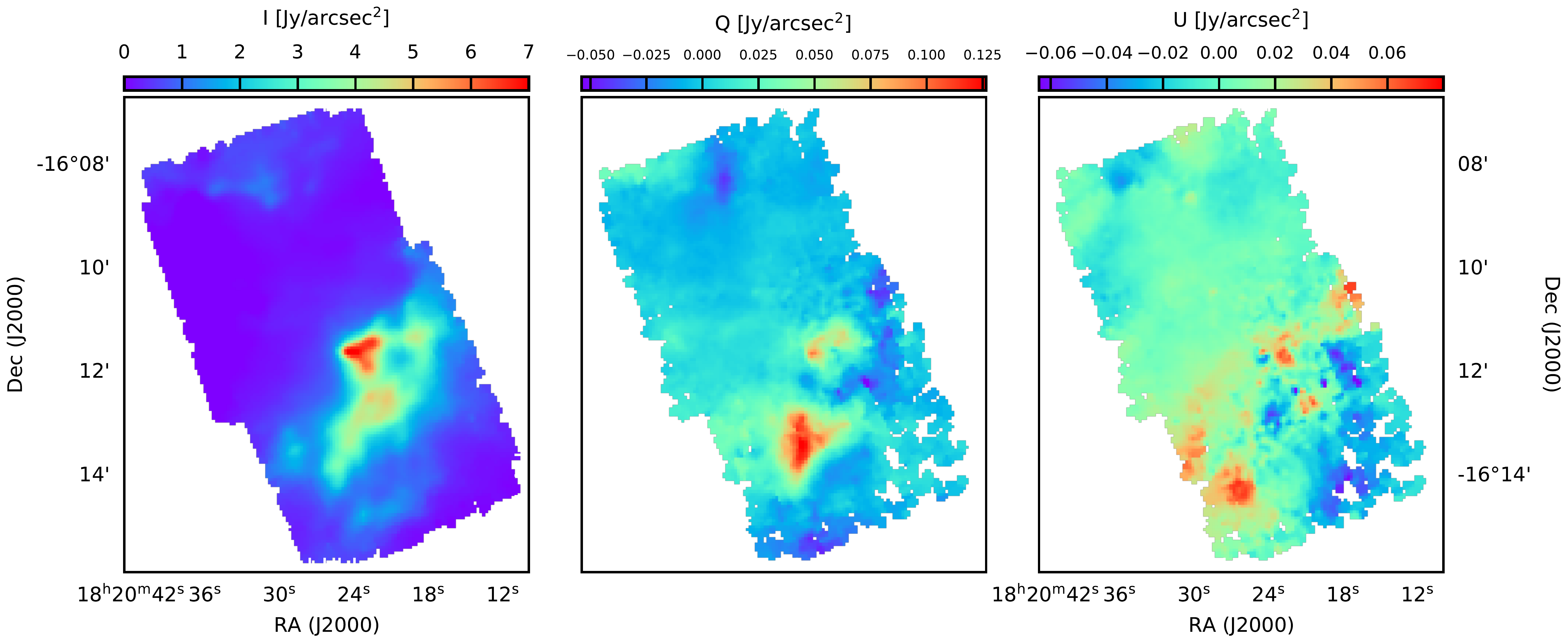}
\caption{From left to right: $I$, $Q$, and $U$ maps.}
 \label{fig2:M17_IQU}
\end{figure*}

\section{Observations}\label{sec:obs}
The thermal dust polarization data of M17 were obtained by SOFIA/HAWC+ instrument centered at \mbox{154 $\rm \mu m$} with a designed beam size of $13.6$\arcsec. The observed region is shown in Figure \ref{fig1:M17_RGB}. The inferred maps have an original pixel-size of $6.9$\arcsec \citep{Harper_2018}. Then, {\it Nyquist} samplings were applied during the data reduction processes to have the final Stokes parameter maps with a re-sampled pixel-size of $\sim 3.4$\arcsec. The Stokes $I$, $Q$, and $U$ maps are shown in Figure \ref{fig2:M17_IQU}.

The biased polarization fraction, $p_{\rm bias}$, is calculated as follows \citep{gordon2018sofia}:
\begin{equation}
    p_{\rm bias}=100 \sqrt{ \left(  \frac{Q}{I} \right)^2 + \left(  \frac{U}{I} \right)^2 } = 100 \frac{I_p}{I} \quad [\%],
\end{equation}
where $ I_p =  \sqrt{Q^2 + U^2}$ is the polarization intensity. The associated error on the biased polarization fraction is 
\begin{equation}
    \sigma_p = p_{\rm bias} \left[  \left( \frac{\sigma_{I_p} }{ I_p} \right)^2   +  \left( \frac{\sigma_{I} }{ I} \right)^2    \right]^{1/2},
\end{equation}
where $ \sigma_{I_p}$ and $ \sigma_I$ are the uncertainties on $ I_p$ and $ I$, respectively. The error propagation on the non-linear function $ I_p (Q, U)$ is expanded as
\begin{equation}
\label{eq:error_propagation}
    \sigma_{I_p}^2 = \left| \frac{ \partial I_p}{\partial Q}     \right|^2 \sigma_Q^2  + \left| \frac{ \partial I_p}{\partial U}     \right|^2 \sigma_U^2  + 2 \frac{\partial I_p }{\partial Q} \frac{\partial I_p }{ \partial U} \sigma_{QU}.
\end{equation}
Assuming $Q$ and $U$ are uncorrelated, namely the covariance term $\sigma_{QU} = 0$, we obtain the error for the polarization intensity

\begin{equation}
    \sigma_{I_p} = \left[ \frac{  Q^2 \sigma_Q^2 + U^2 \sigma_U^2 }{Q^2 + U^2}  \right] ^{1/2},
\end{equation}
where $\sigma_Q$ and $ \sigma_U$ are the errors on Stokes $Q$ and $U$, respectively. The debiased polarization fraction, $p$, is calculated following the approach by \cite{Wardle_1974}
\begin{equation}
    p = \sqrt{p_{\rm bias} ^2 - \sigma_p ^2}.
\end{equation}
The polarization angle, $\theta$, is defined as
\begin{equation}
    \theta = \frac{1}{2} \tan^{-1} \left( \frac{U}{Q} \right),
\end{equation}
and the error on the polarization angle is calculated as follows
\begin{equation}
    \sigma_\theta = \frac{\sqrt{ (Q \sigma_U)^2 + (U \sigma_Q)^2}}{2 (Q^2 + U^2)} .
\end{equation}

A detailed exploration of the raw data is presented in Appendix \ref{appendixA}. In the following, we apply two quality cuts to the data, requiring a high signal-to-noise ratio (SNR) on the measurements of (1) the intensities $\rm SNR_I > 236$ and (2) the polarization fraction $\rm SNR_p > 3$. In what follows, we call this `master cut'. This quality cut automatically satisfies the third criterion recommended by the SOFIA collaboration for high-quality scientific data, namely $p < 50\%$ \citep{gordon2018sofia}. The choice of the cut on $\rm SNR_I$ is to have the measurement uncertainties on the polarization fraction, $\sigma_p$, better than $0.6 \% $. The approximate correlation between these two quantities is ${\rm SNR_I}= \dfrac{\sqrt{2} }{\sigma_p} \approx 236$ \citep{gordon2018sofia}. After applying the master cut, the remaining pixels are 5139, corresponding to 23\% of an original map of the size 138$\times$162 (22356 pixels). The master cut excludes the low level emission and significantly improves the quality of the data. Its performance is illustrated in Figure \ref{fig:M17_SNR_cut} and Table \ref{tab:SNR}. Subsequent analyses are carried out using this master cut.

\begin{figure*}[htbp!]
\includegraphics[width=0.98\textwidth]{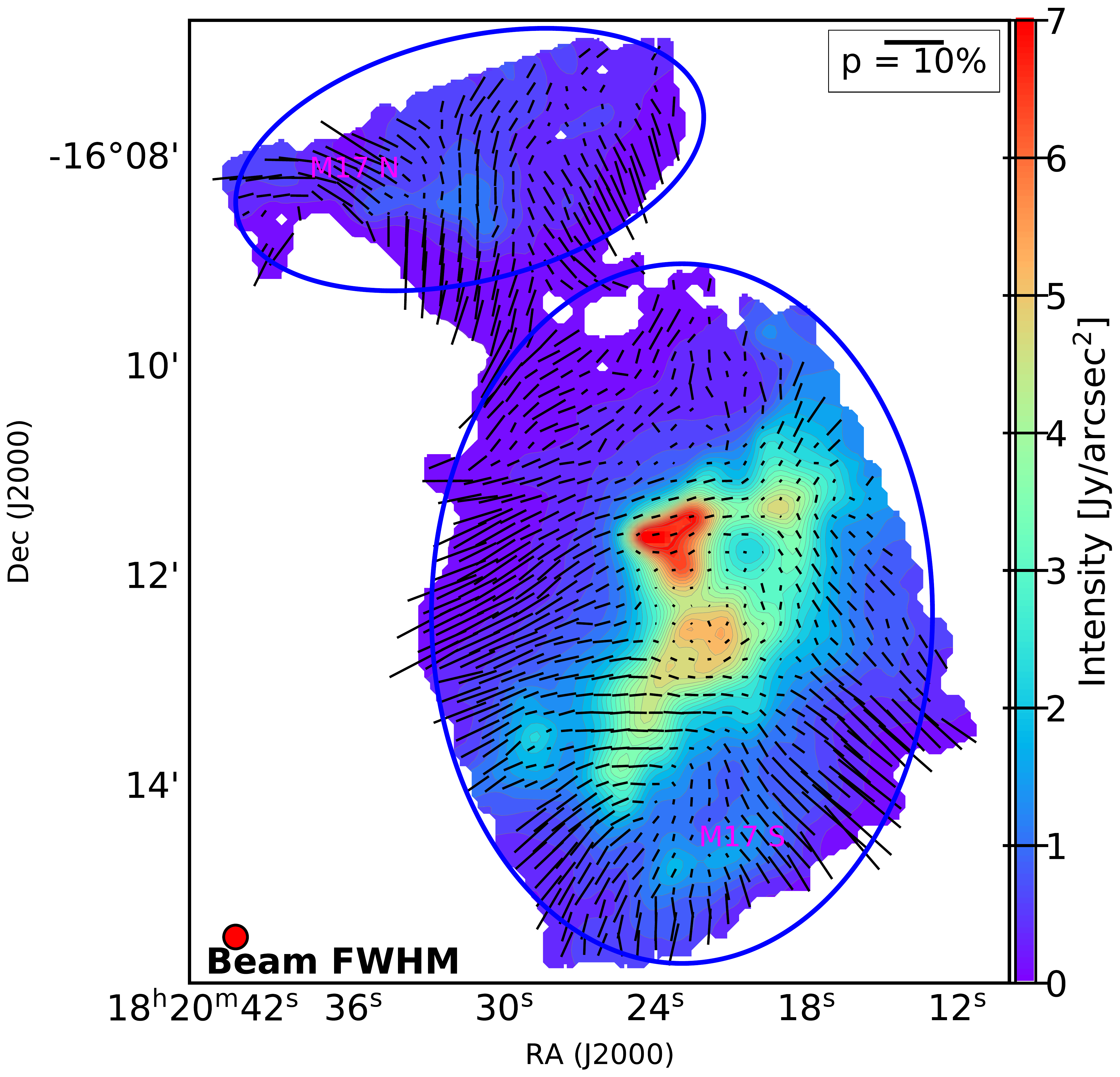}
\caption{The inferred B-field orientation map of the M17 region observed by SOFIA/HAWC+. The half-vectors are proportional to the polarization fraction and rotated by $90^\circ$ from the polarization angles to trace the B-fields. The color scale shows the total intensity in units of $\rm Jy\;arcsec^{-2}$. The intensity map is smoothed with a kernel beam of 3$\times$3 pixels across. A line segment of $10\%$ polarization is shown as a reference in the upper right corner. The beam size is shown in the lower left corner. The blue ellipses are the M17-N and M17-S regions mentioned in the text.}
\label{fig3:M17_pol_fulmap}
\end{figure*}

Figure \ref{fig3:M17_pol_fulmap} shows the inferred B-field orientation map of M17. The lengths of the line segments are proportional to the polarization fraction, and their orientations are obtained by rotating 90$^\circ$ from the polarization angles to trace the B-fields. These segments are called `half-vectors' since the directions of the vectors are not known. The map gives the first impression of the general B-field morphology and the polarization properties of the 154 $\mu$m emission of the region. 
\begin{figure*}[htbp!]
\centering

    \includegraphics[width=0.49\textwidth]{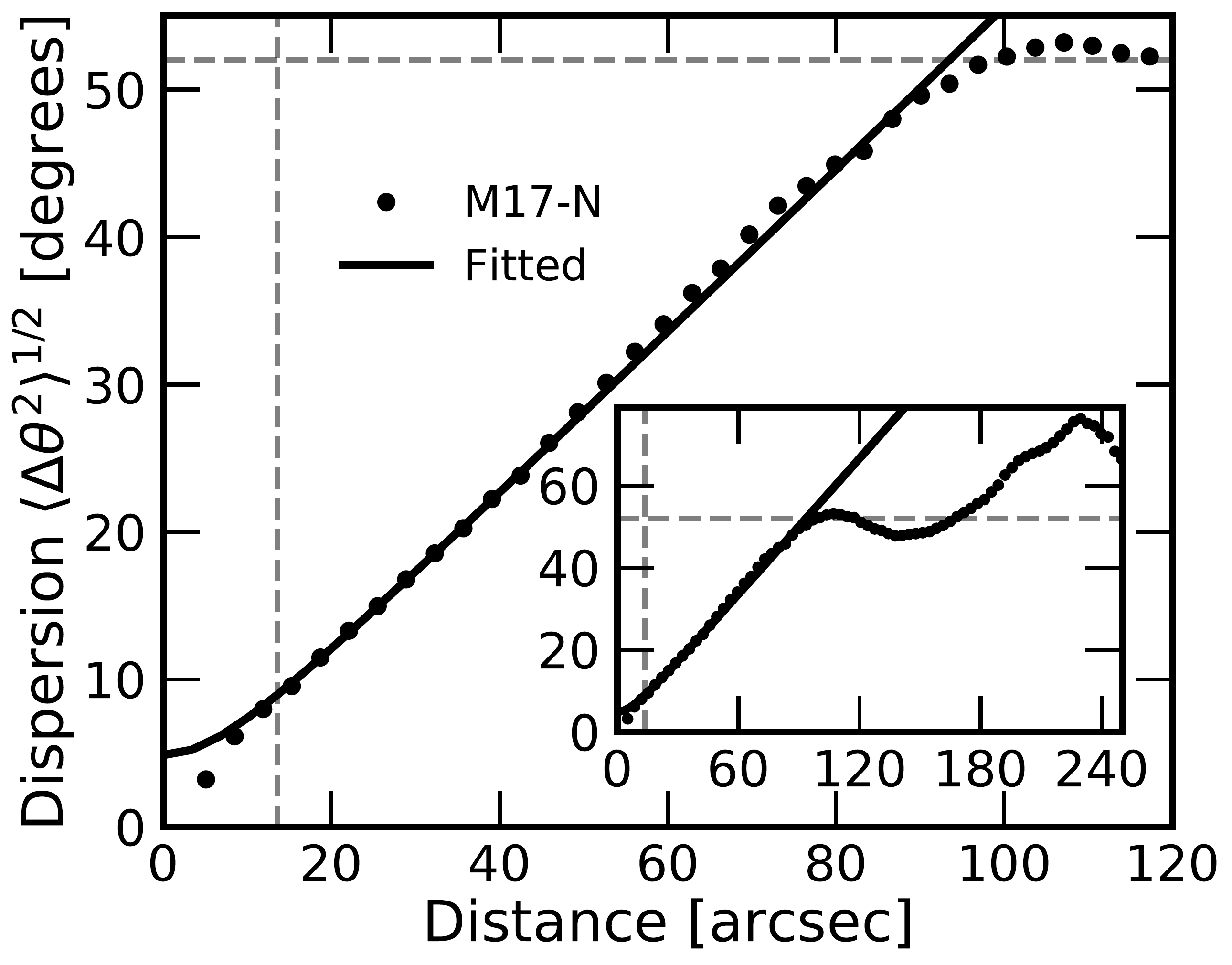}
    \includegraphics[width=0.47\textwidth]{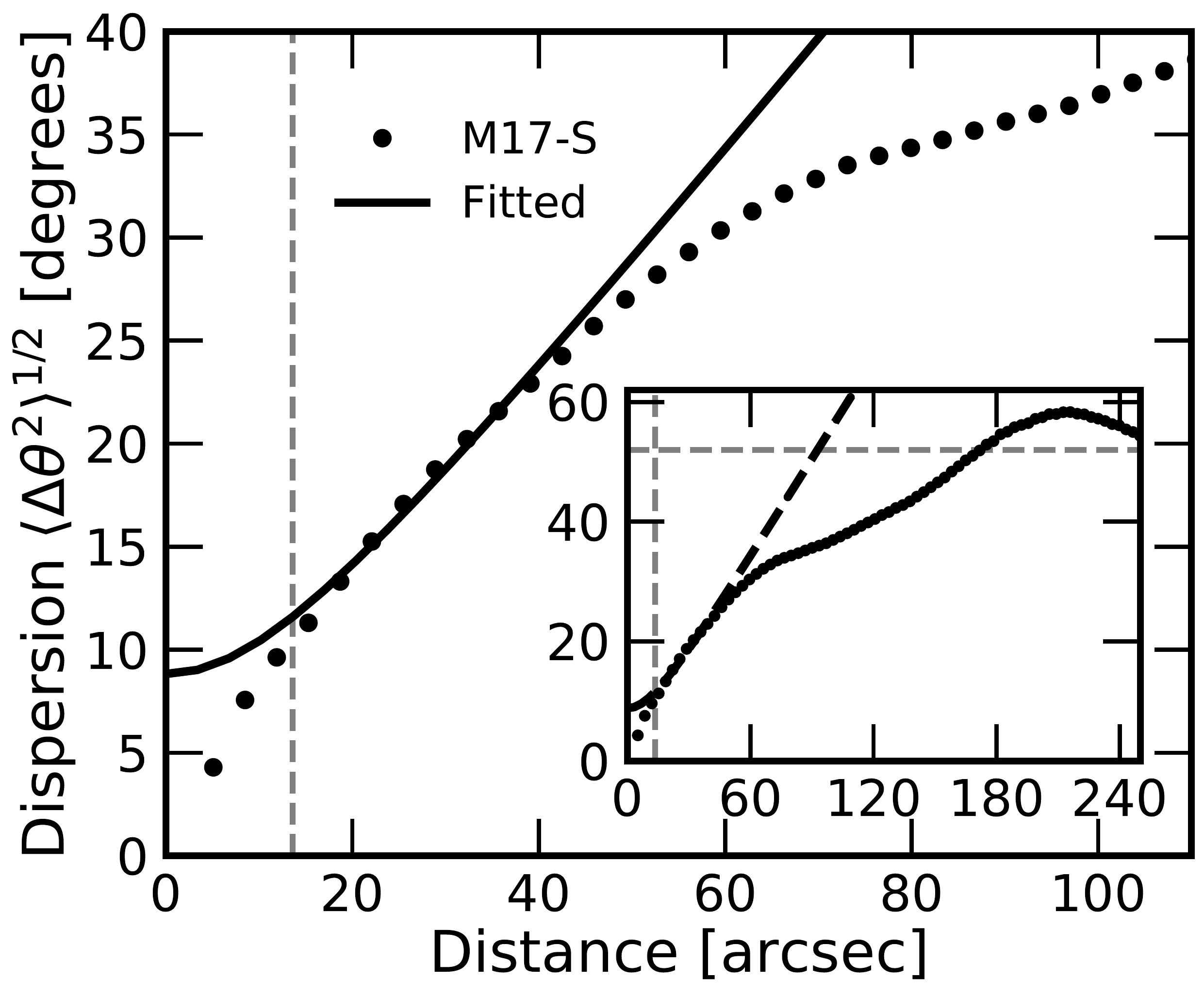}
    
\caption{Structure functions $\langle \Delta \theta^2 (\ell)\rangle ^{1/2}$ for M17-N (left), and M17-S (right): black dot lines are the data and black curves are the fitted models. In the inserts, $\langle \Delta \theta^2 (\ell)\rangle ^{1/2}$ of M17-N and M17-S are shown in black curves for the full range of $\ell$, respectively. The horizontal dashed lines are the random field value of $ \pi / \sqrt{12} \sim 52 ^\circ$ \citep{serkowski1962, planck2015_dust_pol}. The vertical dashed lines are the SOFIA/HAWC+ beam size of 13.6\arcsec at 154 $\rm \mu m$.}
\label{fig4:M17_struc_func}
\end{figure*}

%%%%%%%%%%%%%%%%%%%%%%%%%%%%%%%%%%%%%%%%%%%%%%%
\section{Data analysis}
\label{sec:analysis}
The Davis-Chandrasekhar-Fermi method is one of the most well-known techniques for the estimation of the B-field strength. There is some controversy around the method \citep[e.g.,][]{Pattle_2019_review, liu2021calibrating}, however the DCF method still provides a mean for estimating the magnetic field strength from dust polarization measurements. The method is based on the fact that turbulent motions induce a turbulent component on top of the mean, large-scale magnetic field and the assumption that the turbulent kinetic energy is balanced by the induced magnetic energy. The strength of B-fields in the plane-of-sky, $B _{\rm POS}$, is calculated using the following equation \citep{crutcher_2004}
\begin{equation}
\label{eq:B_mag}
    B_{\rm POS} = Q_c \sqrt{4 \pi \rho} \frac{\sigma_\nu}{\sigma_{\theta} } \approx 9.3 \sqrt{ n(\rm H_2) } \frac{\Delta V}{ \sigma_{\theta}  } \ [\rm \mu G],
\end{equation}
where $ Q_c$ is a factor of the order of unity to correct for the light-of-sight and beam-integration effects, $ \rho$ is the gas density in $\rm g\;cm^{-3}$, $\Delta V = \sigma_\nu \sqrt{8 \ln{2}} = 2.355 \sigma_\nu$ is the FWHM of the one dimensional non-thermal velocity dispersion, $\sigma_\nu$, in $\rm km\;s^{-1}$, $n(\rm H_2)$ is the volume density of the molecular hydrogen in $\rm cm^{-3}$, and $\sigma_\theta$ is the polarization angle dispersion in degrees. In general, the magnetic field strength in MCs is in the range of micro- to milligauss.

Figure \ref{fig3:M17_pol_fulmap} indicates two distinct regions encompassed by the two ellipses in terms of physical conditions, including density, temperature, and dynamics, as can be seen in Figures \ref{fig5:M17_CO_moment0_1}, \ref{fig7:M17_NH2}, and \ref{fig10:M17_temp} in the next sections. The  ellipse in the north has a center at RA $\sim$ $\rm 18^h20^m31^s.6$ and DEC $\sim$ $\rm-16^\circ08\arcmin 04\arcsec.3$, major $\times$ minor axes of 137\arcsec $\times$ 69\arcsec, and a position angle of $14^\circ$. The ellipse in the south has a center at RA $\sim$ $\rm 18^h20^m23^s.2$ and DEC $\sim$ $\rm-16^\circ12\arcmin 24\arcsec$, major $\times$ minor axes of 200\arcsec $\times$ 143\arcsec, and a position angle of $0^\circ$. Therefore, we divide the map into two corresponding northern and southern regions coined M17-N and M17-S for further analyses. Subsequently, we are going to identify the input parameters for the DCF method to calculate the B-field strengths of the two regions.

\subsection{Polarization angle dispersion: Structure function}
Structure function method was initially proposed by \citet{Falceta_Gon_alves_2008} and \citet{Hildebrand_2009} for evaluating the polarization angle dispersion, $\sigma_\theta$, in the plane-of-sky. Fundamentally, this method calculates a two-point correlation function for pairs of the polarization angles as follows,
\begin{equation}
\label{eq:struc_func}
 \langle \Delta \theta \left( \ell \right) \rangle^ {1/2} =
  \Bigg\{  \frac{1}{N \left( \ell \right)} \sum _{i=1}^{N \left( \ell \right)}     \left[   \theta \left( \mathbf{x} \right)   - \theta \left(  \mathbf{x} + \Bell
 \right)  \right]^2 \Bigg\}^{1/2},
\end{equation}
where $\langle ... \rangle$ denotes an average; $N(\ell)$ is the number of pairs of pixels having a displacement between the two pixels $\lvert \Bell \rvert = \ell$; $\mathbf{x}$ and $ \mathbf{x} + \Bell$ are the location vectors of the two considered pixels with corresponding polarization angles $ \theta (\mathbf{x})$ and $ \theta ( \mathbf{x} + \Bell)$, respectively. We note that only the pairs with $\Delta \theta (\ell) = \left| \theta (\mathbf{x} ) -  \theta (\mathbf{x} + \Bell) \right| < 90^\circ$ are retained for further analyses because the directions of the half-vectors are unknown. Assuming that the B-fields have two independent components (a large-scale structure field, $B_0$, and a turbulent one, $\delta B$), the structure function can be written as below for a small displacement $\ell$ \citep{Hildebrand_2009, crutcher_richard}
 \begin{equation}
 \label{eq9}
     \langle \Delta \theta \left( \ell \right) \rangle = b^2 + m^2\ell^2 + \sigma^2_M (\ell),
 \end{equation}
 where $b$ is the contribution of turbulence, $m \ell$ of the large-scale structure component; and $\sigma_M (\ell)$ is the measurement uncertainties calculated for each pair of polarization angles. In practice, $\sigma_M (\ell)$ is taken into account while calculating $\langle \Delta \theta \left( \ell \right) \rangle^ {1/2}$ in Equation \ref{eq:struc_func}. The ratio of the turbulent to large-scale structure components is expressed as \citep{Hildebrand_2009}
 \begin{equation}
    \label{eq:eq_ratio}
     \frac{\delta B}{B_0 } = \dfrac{b}{ \sqrt{ 2 - b^2} }.
\end{equation}

From the definition of polarization angle dispersion, $\sigma_\theta$, in Equation \ref{eq:struc_func} combining with Equation \ref{eq9} neglecting the contribution of the large-scale structure component, $m \ell$, we have $\sigma_\theta = b/\sqrt{2}$.

We calculate the structure functions for M17-N and M17-S separately. In general, they display similar rising tendencies at small displacements $\ell$ and reach the value of random field of $52^\circ$ at large displacements (see Figure \ref{fig4:M17_struc_func}). The extent of M17-N is smaller than that of M17-S, hence the maximum displacement of M17-N is smaller.
We then fit the calculated structure functions with the $\ell$-dependence of the polarization angle dispersion using Equation \ref{eq9}. The fits were carried out for $\ell > 13.6\arcsec$ to avoid the beam effects. The upper $\ell$-limits were chosen to achieve the best fits (black curves in Figure \ref{fig4:M17_struc_func}). Table \ref{tab1:summary} lists the obtained values $b$ = $4.9 \pm 0.2$ and $8.8 \pm 0.7 $ which means \mbox{$\sigma_{\theta}$ = $3.5^\circ \pm 0.2^\circ$} and $6.2^\circ \pm 0.5^\circ $ for M17-N and M17-S, respectively. The uncertainties on $b$ are from the fits. We note these values are smaller than that found by \citet{Hildebrand_2009}, $\sigma_{\theta} \sim 10^\circ$, using data from Caltech Submillimeter Observatory at $350 \ \rm \mu m$ with a spatial resolution of $\sim 20\arcsec$ for M17. Table \ref{tab1:summary} also lists the ratios of the turbulent to large-scale structure B-field components calculated using Equation \ref{eq:eq_ratio} which shows that the turbulent fields are much smaller than the large-scale one in M17.

\subsection{Velocity dispersion} 
\label{sub:velocity}
\begin{figure*}[!thbp]
\includegraphics[width=1.0 \textwidth]{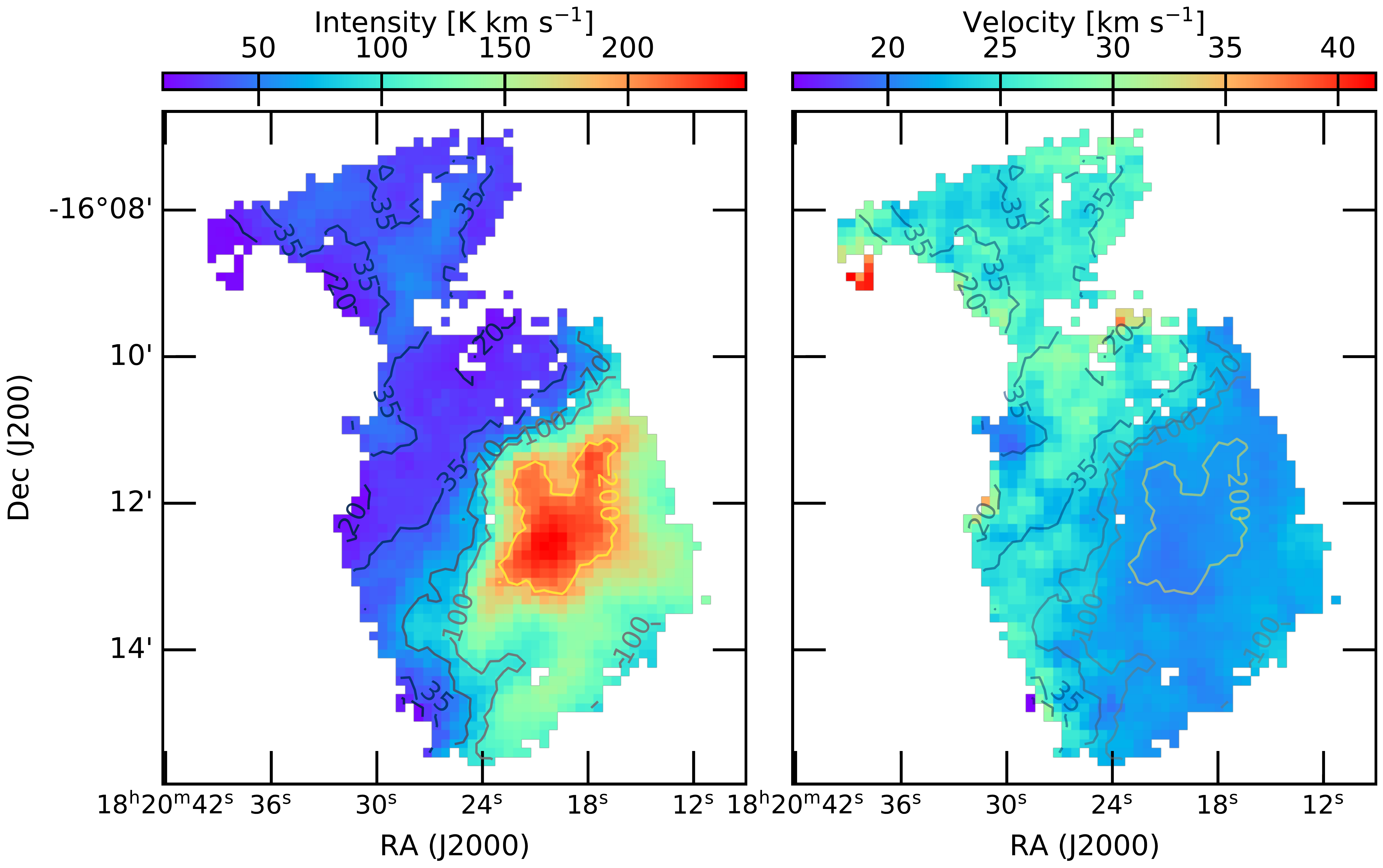}
\caption{Left: velocity-integrated intensity map. Right: mean velocity map. The contour levels are at 20, 35, 70, 100, and 200 K\;km\;s$^{-1}$ for both maps.}
\label{fig5:M17_CO_moment0_1}
\end{figure*}

\begin{figure*}[!thbp]
\includegraphics[width=1.0\textwidth]{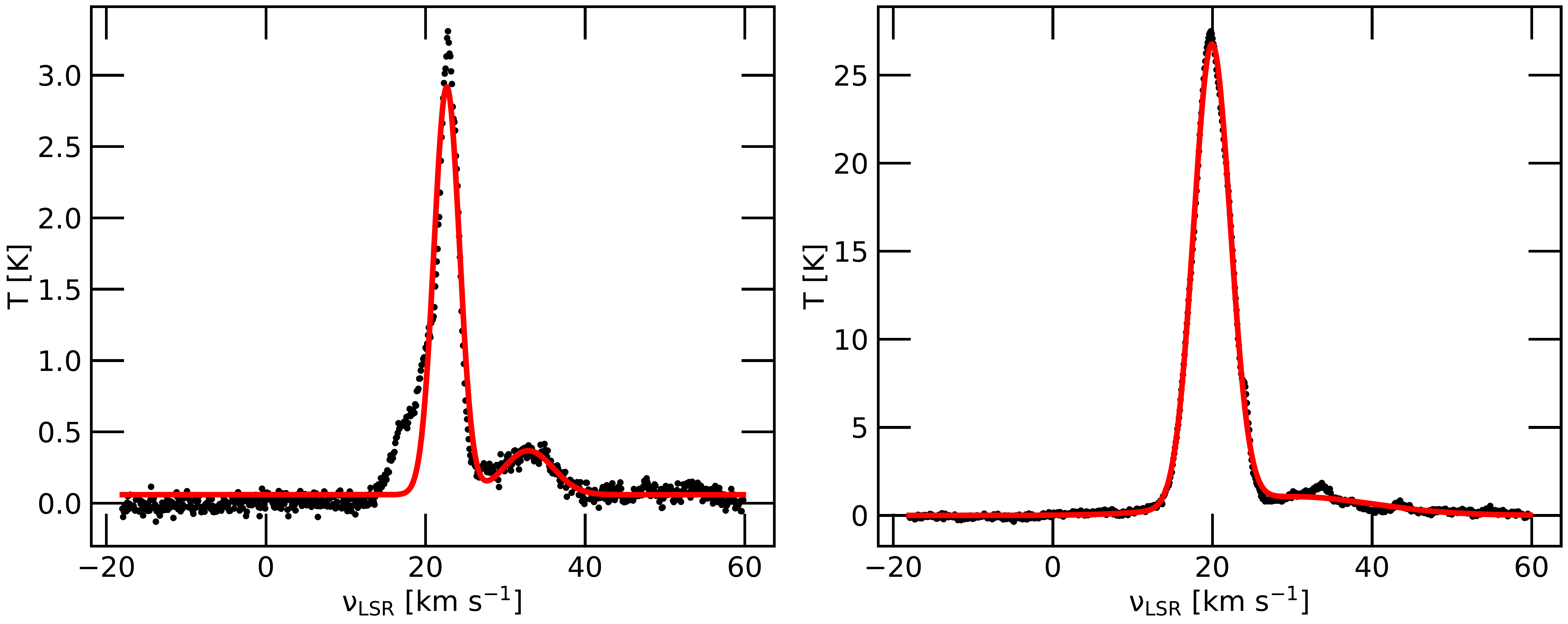}
\caption{Average integrated spectra (black) with two-Gaussian fits (red) for M17-N (left) and M17-S (right).}
\label{fig6:M17_CO_velo}
\end{figure*}

We estimate the velocity dispersion of M17 using the publicly available archival data $\rm ^{13}CO \ (J=1 \rightarrow 0)$ taken by Nobeyama 45-m telescope \citep{Nakamura_2019}. The measurements were made at the central frequency $\nu_{\rm central} \sim 110.201 \ \rm{ GHz}$ with a spectral resolution of \mbox{0.1 km\;s$^{-1}$}. The resulting map has a pixel size of 7.5\arcsec and a beam size of 14.9\arcsec which was then cropped to match with the M17 observed region by SOFIA/HAWC+. Figure \ref{fig5:M17_CO_moment0_1} presents the velocity-integrated intensity (left) and the mean velocity (right) maps. The integration was done over the entire velocity range from $ -20$ to $ 60 \ \rm km \; s^{-1}$ in the local standard of rest (LSR) frame. Figure \ref{fig6:M17_CO_velo} displays the integrated spectra averaged over the number of pixels for M17-N (left) and M17-S (right). They can be well-described by two-Gaussian fits giving the mean positions and corresponding standard deviations, $\sigma_{\nu;{\rm ^{13}CO}}$, of the peaks. The spectra of M17-N and M17-S show the main peaks at $22.7 \pm 1.6$ and $19.9 \pm 2.3$ km\;s$^{-1}$, respectively. These main peaks are much higher than the second ones. Therefore, we only use the velocity dispersion of these two peaks for the calculation of B-field strengths in M17-N and M17-S. Table \ref{tab1:summary} summarizes the results of the fits which are in agreement with the results obtained by \citet{Nakamura_2019, Nguyen_Luong_2020}, although their results are for wider areas of M17. The statistical uncertainties of the velocity dispersion from the two-Gaussian fits are smaller than the spectral resolution of Nobeyama, therefore, we take the uncertainties equal to the spectral resolution of 0.1 km\;s$^{-1}$. We note that M17-N is affected by the outflows and shocks from G015.128\footnote{G015.128 is a massive young stellar object (MYSO) and likely an A-type super-giant star \citep{Pomohaci2017}. The current SOFIA/HAWC+ data does not fully cover this source.} \citep{Lim_2020} which is expected to be more turbulent than M17-S. 
\begin{figure*}[!htbp]
\centering
\includegraphics[width=1.0\textwidth]{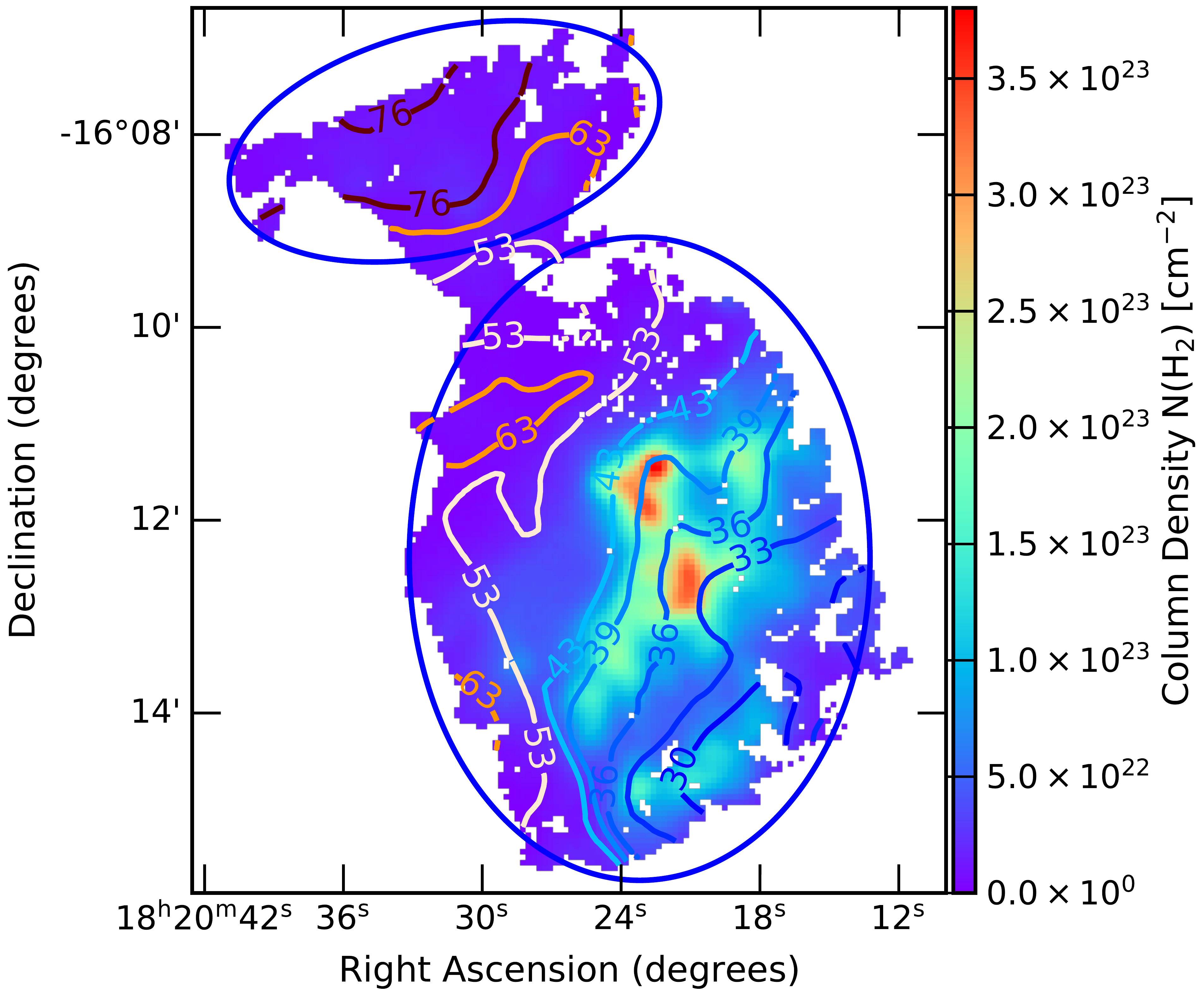}
\caption{Map of the column density, $ N(\rm H_2)$, of M17. The contours display the dust temperature, $T_{\rm d}$, values as shown in Figure \ref{fig10:M17_temp}. The blue ellipses are the same as in Figure \ref{fig3:M17_pol_fulmap}. In M17-S, the relation of $T_{\rm d}$ and $N(\rm H_{2})$ is complex. $T_{\rm d}$ is positively correlated to $N(\rm H_{2})$ up to $\sim 43\, \rm K$, but turns to negatively otherwise.}
\label{fig7:M17_NH2}
\end{figure*}

\begin{table*}[!htbp]
\centering
\caption{Summary of the obtained results. $b$ is the rms contribution of the turbulent component and
$m$ of the large-scale B-fields to the polarization angle dispersion, $\sigma_{\theta}$. $\delta B/B_0$ is the ratio of the turbulent to the large-scale fields. Detailed description of the parameters can be found in the text.}
\label{tab1:summary}
\begin{tabular}{|c|c|c|c|} 
\hline
\hline
\multicolumn{2}{|c|}{ Parameters} & M17-N & M17-S             \\ 
\hline
\multicolumn{2}{|c|}{$b$} & $ 4.9 \pm 0.2 $   & $ 8.8 \pm 0.7 $ \\
\multicolumn{2}{|c|}{$m$} & $ 0.55 \pm 0.002$ & $ 0.55 \pm 0.01$ \\
\multicolumn{2}{|c|}{Polarization angle dispersion, $\sigma_{\theta}$ [deg] }   & $ 3.5 \pm 0.2 $ & $ 6.2 \pm 0.5 $ \\
\multicolumn{2}{|c|}{$\delta B/B_0 $} & 0.06  & 0.11 \\
\hline
\multicolumn{2}{|c|}{Peak position, $ \nu_{\rm LSR} \ \rm [km\;s^{-1}]$}   &  22.7 $\pm$ 0.1 &  19.9 $\pm$ 0.1   \\
\multicolumn{2}{|c|}{Velocity dispersion, $ \sigma_{\nu,{\rm ^{13}CO }} \ \rm [km\;s^{-1}]$}  &  1.6  $\pm$ 0.1 &    2.3  $\pm$ 0.1   \\
\hline
\multicolumn{2}{|c|}{Column density, $ \langle N({\rm H_2}) \rangle \ \rm [cm^{-2}]$} &  $(9.1\pm 4.0) \times 10^{21} $  & $(6.2\pm 6.5) \times 10^{22} $ \\
\hline
\multicolumn{2}{|c|}{Volume density, $n({\rm H_2}) \  \rm [cm^{-3}]$ } &  $ (9.3\pm 4.1) \times  10^{3}$  &  $  (4.2 \pm 4.4) \times 10^{4}$\\
\hline
\end{tabular}
\end{table*}

The obtained standard deviations are then converted to the non-thermal velocity dispersion using $ \sigma^2_{\nu} = \sigma^2_{ \nu;{\rm ^{13}CO}} - \dfrac{k_{\rm B} T}{m_{^{13}\rm CO } }$, where $m_{ ^{13}\rm CO }$ is the $\rm ^{13}CO$ molecule mass equal to 29 amu, $k_{\rm B}$ is the Boltzmann constant, and $T$ is the gas temperature. It turns out that the thermal contribution to the velocity dispersion is negligible if we adopt the average gas temperature of 20 K according to \citet{Nguyen_Luong_2020}. Therefore, we take the non-thermal velocity dispersion for the M17-N and M17-S regions equal to the standard deviations of their corresponding main peaks listed in Table \ref{tab1:summary}.

\subsection{Column and volume densities}\label{sec:nh}
The column density, $N({\rm H_2})$, has been derived based on a graybody (i.e., modified blackbody) fit with the filter weighted opacity, $\kappa_{\rm \nu}$, and blackbody radiation, $B_{\nu}(T_{\rm d})$. %where $T_{\rm d}$ is the dust temperature. Comment-TH: Td already defined previously
As following the techniques described in \citet{Lim_2016}, {\it Herschel} 160, 250, and $350\;\micron$ data of M17 have been convolved to the beam size of {\it Herschel} $500\;\micron$ images ($\sim 36\arcsec$) before executing the pixel-by-pixel graybody fitting via all four band data. The template $T_{\rm d}$ of angular resolution $\sim 36\arcsec$ is re-gridded to match to the pixel geometry of SCUBA2 $850\;\micron$ map (angular resolution $\sim 14 \arcsec$ and pixel size=$4\arcsec$). We then repeated the graybody fit by utilizing SCUBA2 $850\;\micron$ map \citep{Reid_2016a} of M17 with the re-gridded $T_{\rm d}$ map. Figure \ref{fig7:M17_NH2} shows the resulting $N({\rm H_2})$ map superimposed by the dust temperature (Figure \ref{fig10:M17_temp}). Using this map, we calculate the total column densities, $N_{\rm total, H_2}$, the average column densities, $\langle N({\rm H_2}) \rangle$, as well as their associated uncertainties, $\sigma_{N({\rm H_2})}$, for the M17-N and M17-S regions. The results are presented in Table \ref{tab1:summary}.

We then follow the approach described in \citet{Lee_2012, li2014, ngoc2020observations} to estimate the volume density, $ n(\rm H_2)$. The mass of a region is calculated by $ M = \beta m_{\rm H_2} N_{\rm total,H_2} (D \Delta)^2 $, where $ \beta = 1.39$ accounts for $\rm He$ of $10\%$ abundance in addition to $\rm H_2$ in the total mass, $m_{\rm H_2}$ is the mass of hydrogen molecules, 
%$N_{\rm total,H_2}$ is the total column density estimated from the $ N({\rm H_2})$ map, 
%Comment-TH: NH2 already defined, no need to explain again
$ D = 1.98 \ \rm kpc $ is the distance to M17 \citep{Xu_2011}, and $\Delta = 4\arcsec$ is the pixel size of the $N({\rm H_2})$ map. Then the volume density $n(\rm H_2)$ is expressed as
\begin{equation}
\label{eq:volume_density}
   n ({\rm H_2}) = \frac{3M}{ 4 \pi R^3 m_{\rm H_2} } = \frac{3 \beta  N_{\rm total,H_2}  (D \Delta)^2 }{ 4 \pi R^3} \quad  \rm [cm^{-3}],
\end{equation}
where $R$ is the radius of a considered clump equal to $R = \sqrt{(ab)/2}$ ($a$ and $b$ are the major and minor axes of an ellipse covering the clump \citep{li2014}). Equation \ref{eq:volume_density} is deduced assuming spherical geometry of the molecular clump. As shown in Figure \ref{fig3:M17_pol_fulmap} and \ref{fig7:M17_NH2}, we define two ellipses that encompass the observed M17-N and M17-S regions with corresponding angular radii of $69\arcsec$ and $120\arcsec$, respectively. The physical radii of the two clumps are obtained by multiplying $R$ with the distance $D$ to M17. Employing Equation \ref{eq:volume_density}, the volume densities, $n(\rm H_2)$, are calculated, and the results are shown in Table \ref{tab1:summary} for both M17-N and M17-S regions. The uncertainty on $n(\rm H_2)$ is calculated assuming that its relative uncertainty is the same as that of $N_{\rm total,H_2}$, namely $\dfrac{\sigma_{n( \rm H_2)}}  {n(\rm H_2)}  = \dfrac{ \sigma_{N(\rm H_2)} }{ \langle N (\rm H_2)  \rangle  }$.

\section{Results and Discussion}
\label{subsec:result}
In this section, we first describe the magnetic field morphology of the observed region and report the magnetic strengths in the plane of the sky, $B_{\rm POS}$, using the DCF method for the M17-N and M17-S regions. We then explore the relative contribution of B-fields, gravity, and turbulence by calculating the mass-to-flux ratios and Alfv\'enic Mach numbers. Finally, we investigate dust grain alignment and disruption based on the polarization fraction in the observed regions.

\subsection{Magnetic field morphology}
\label{subsec:bfieldmorph}
\begin{figure*}[htbp!]
\label{fig:M17_pol_vector}
\centering
\includegraphics[width=1.0 \textwidth]{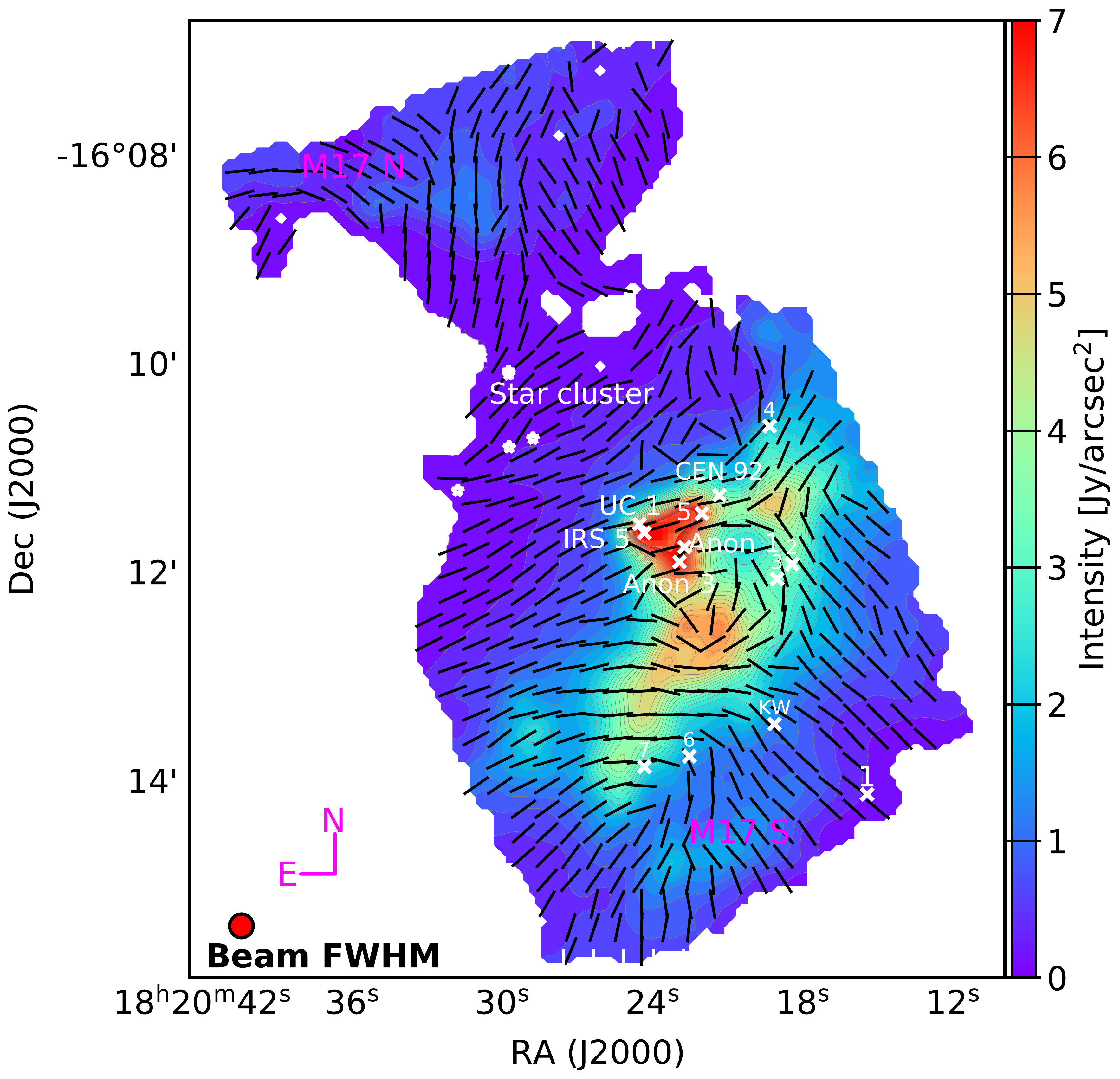}
\caption{The B-field morphology of M17 superimposed on the intensity map. Half-vectors are rotated by $90^\circ$ from the polarization angles to represent the magnetic field orientations. Here, the half-vectors are plotted with uniform length. The star symbols represent the most massive ($>$O7) stars in the open cluster NGC 6618 \citep{hanson1997}. There are several infrared sources present in M17-S such as  UC 1, IRS 5, CEN 92, Anon 1, and Anon 3. The KW (Kleinmann-Wright) object is a binary-star system \cite{KW_object1973}. The locations of the sources are adopted from \citet{Lim_2020}.}
\label{fig8:M17_halfvector}
\end{figure*}

\begin{figure*}[htbp!]
\includegraphics[width=1.0\textwidth]{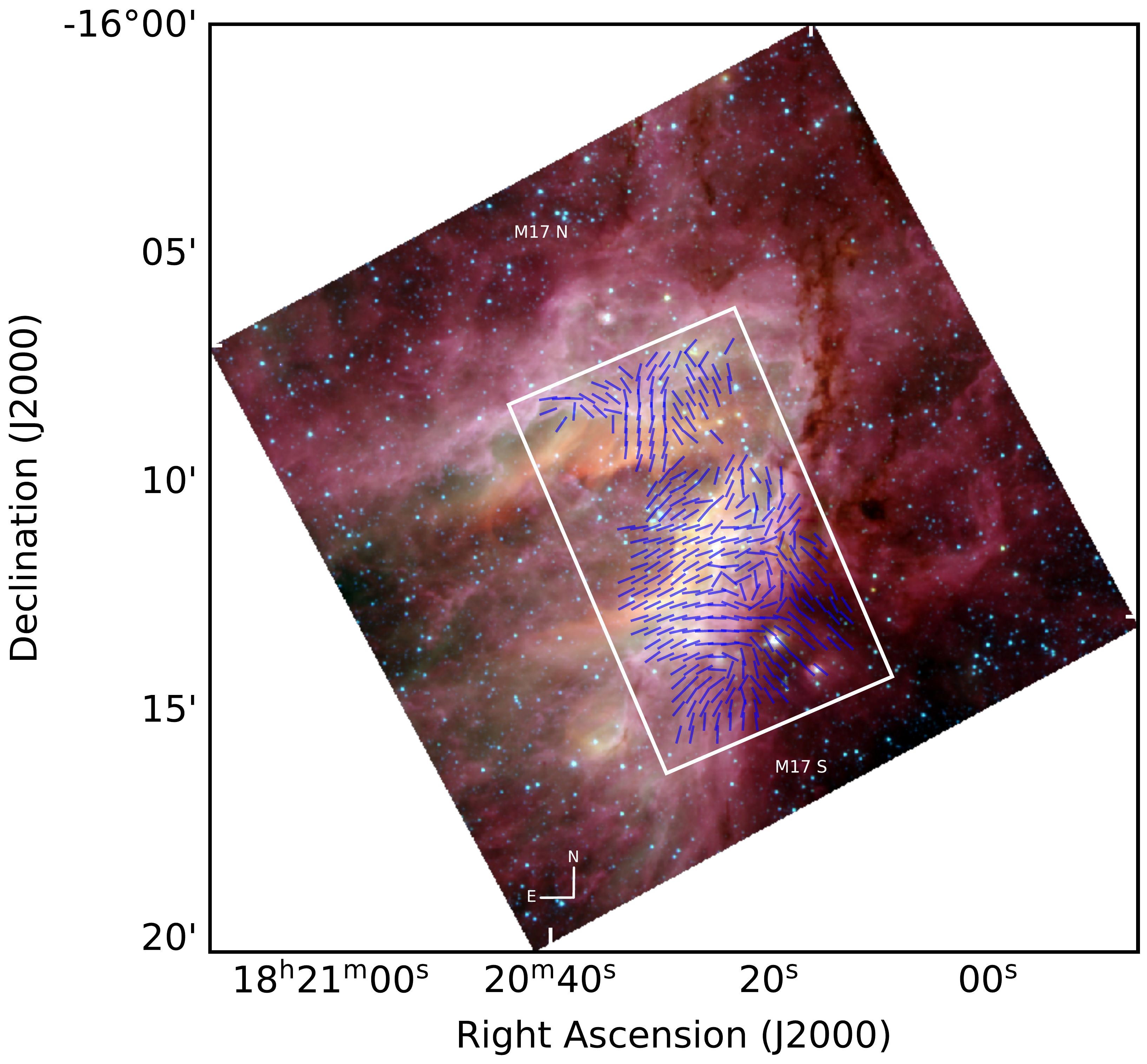}
\caption{A RGB image of M17 using {\it Spitzer} data as described in Figure \ref{fig1:M17_RGB}. The white box is the observed area of SOFIA/HAWC+ toward M17. The half-vectors are plotted with uniform lengths using SOFIA/HAWC+ data and rotated by 90$^\circ$ to trace the magnetic fields. In this figure, we can see the larger scale structure mentioned in the text connected to the M17-N region.}
\label{fig9:M17_rgb_morpho}
\end{figure*}

\begin{figure*}[htbp!]
\includegraphics[width=1.0\textwidth]{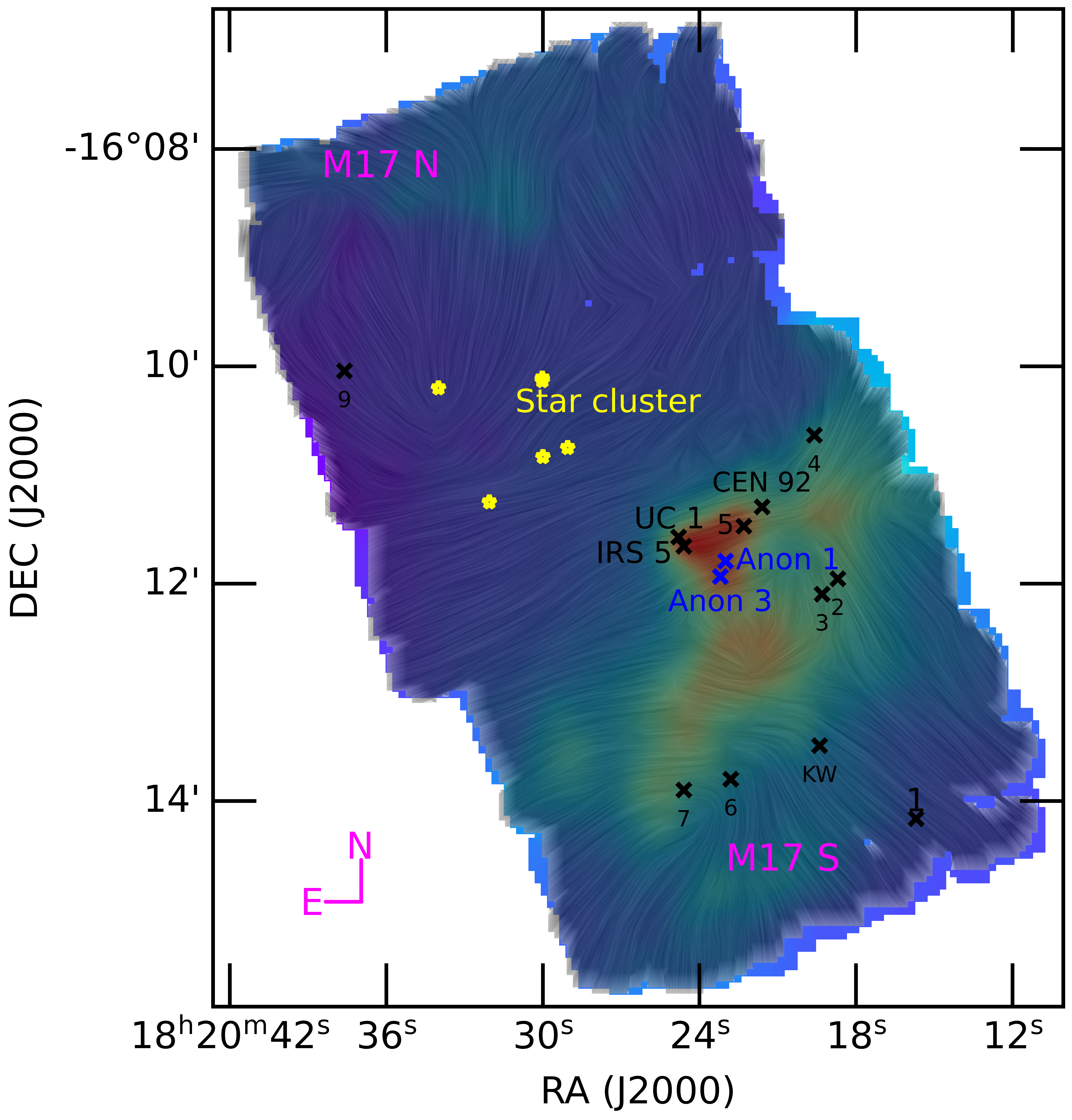}
\caption{Same as Figure \ref{fig8:M17_halfvector} with the `drapery' patterns produced by the line integral convolution (LIC) tool \citep{Cabral_lic}.}
\label{fig9:M17_steplic}
\end{figure*}

The magnetic field morphology is imprinted on star formation processes. We thus can use the observational data to test theoretical predictions of star formation.

Overall, over the whole region, B-fields in the outer, low-density regions are perpendicular to the density structure. Figure \ref{fig3:M17_pol_fulmap} and its close-up version (Figure \ref{fig:a3-closeup}) illustrate well this point with many half-vectors in the low-density regions orthogonal to the intensity contours. Another prominent feature is that when the field lines pass through the \ion{H}{2} region, east of the observed region (see Figure 1 of \citet{Povich_2009} for a more precise location of the \ion{H}{2} region), they tend to run parallel threading M17-N and M17-S. However, in M17-S, the fields run perpendicular to these fields behind the high-density areas at the center (see Figure \ref{fig8:M17_halfvector}).

In the M17-N region, the main B-fields line up along the north-south direction through the highest density area and curve in the eastern and western sides of the region (see Figures \ref{fig8:M17_halfvector} and \ref{fig:a3-closeup}). The morphology of the B-fields in M17-N mimics the fields of a gravitationally collapsed molecular core with the presence of strong magnetic fields where we see the classical hourglass geometry with straight fields in the middle and curved ones at the two wings. This is a common structure which we often find in observations and simulations \citep[see e.g.,][]{Kandori_2018, Wurster_2018,pattle2019}. On the other hand, M17-N is only a small part of a much larger M17-N region (see Figure \ref{fig9:M17_rgb_morpho}) which is not fully covered by the current SOFIA/HAWC+ observations. Therefore, the observed structure may be just a coincidence with the fields following the emission by Polycyclic Aromatic Hydrocarbons (PAHs) at 8 $\mu$m wavelength observed by {\it Spitzer} (the bright pink color structure in the northern region in Figure \ref{fig9:M17_rgb_morpho}). Distinguishing these two scenarios requires observations of higher angular resolution and larger areas of M17-N.

In the M17-S region, generally, the fields run perpendicular to the elongation of the high-density structure. In the region, the fields also form an asymmetric large-scale hourglass shape (see Figures \ref{fig8:M17_halfvector}, \ref{fig9:M17_rgb_morpho}, and \ref{fig9:M17_steplic}). In the southern side of the hourglass, just below Anon 1 and Anon 3, the fields run from east toward west while gradually bending the to south. The fields are highly curved in the most southern part of the region and perturbed in the high density regions at the center. On the northern side of the hourglass, just above UC 1 and CEN 92, we see the curved fields (see Figures \ref{fig8:M17_halfvector} and \ref{fig9:M17_steplic}). This hourglass structure seems to be clearly caused by the gravitational contraction of the massive cores at the center of M17-S. Moreover, the asymmetry is due to the complex distribution of matter in the center and north-western part of the region. Another prominent feature of M17-S, which is commonly found in PDR regions (see e.g., \cite{Pattle_2018}), is the pillar structure. Here, we find a `triangular' pillar with the top-end coincident with the positions of UC 1 and IRS 5 and a `base' in the western direction (see Figures \ref{fig8:M17_halfvector} and \ref{fig9:M17_steplic}). It is interesting to note that the magnetic field morphology found here is similar to the ones reported by \citet{Pattle_2018} where the fields run parallel to the pillar in the \ion{H}{2} region side and perpendicular to the pillar on and behind the base.

\citet{Dotson1996} reported that magnetic field lines are elongated into the cloud core and bulged away from the \ion{H}{2} region, heated by OB stars. This could be an evidence that the \ion{H}{2} region is expanding into its surrounding medium \citep{Zeng_2013}. As a conclusion, our magnetic field morphology is consistent with the scenario that the magnetic field is distorted by \ion{H}{2} region (see more in Section \ref{sec:feedback}), several IR sources, and stellar clusters.

\subsection{Magnetic field strengths}
Given the estimated values of the polarization angle dispersion, $\sigma_\theta$, the one dimensional non-thermal velocity dispersion, $\sigma_\nu$, and the volume densities, $n(\rm H_2)$, in Table \ref{tab1:summary}, we calculate the strengths of B-fields in the plane of the sky using Equation \ref{eq:B_mag}, yielding $B_{\rm POS}=980 \pm 230 \ \rm \mu G$ and $1665 \pm 885 \ \rm \mu G$ for M17-N and M17-S, respectively (see more details in Table \ref{tab:Bpos}). 
The magnetic field strength of M17-S is greater than that of M17-N because of its higher density. 

We note that the DCF method is applicable here since all the polarization angle dispersion is smaller than 25$^\circ$ \citep{crutcher_2004}. The uncertainties are propagated from the uncertainties on $\sigma_\theta$, $\sigma_\nu$, and $ n(\rm H_2)$ using the following equation
\begin{equation}
\label{eq:Bpos_uncer} 
\frac{\delta B_{\rm POS}}{ B_{\rm POS}} = \sqrt{ \left( \frac{1}{2} \frac{ \delta n(\rm H_2) }{ n(\rm H_2) } \right)^2   + \left( \frac{\delta \Delta V}{ \Delta V} \right)^2  +  \left( \frac{\delta \sigma_\theta }{\sigma_\theta} \right)^2 },
\end{equation}
where $\delta n(\rm H_2), \ \delta \Delta V$, and $ \delta \sigma_\theta $ are the uncertainties on $\sigma_\nu$, $\Delta V = 2.355 \sigma_\nu$, and $\sigma_{\theta}$, respectively.
\begin{table}[htbp!]
\centering
\caption{Results of the magnetic field strengths, Alfv\'enic Mach numbers, mass-to-flux ratios for M17-N and M17-S. $ \gamma$ is the inclination angle of the B-fields with respect to the line of sight.}
\label{tab:Bpos}
\begin{tabular}{|c|c|c|c|} 
\hline
\hline
   Region   & $ B_{\rm POS}$    & Mach number  & Mass-to-flux ratio \\
      & $ [\rm \mu G]$     &  $ \mathcal{M_A}$  &  $\lambda$  \\ 
\hline
M17-N &  $980 \pm  230$   &   $0.12 \sin \gamma$   &  0.07 $\pm$ 0.04   \\
\hline
M17-S &  $1665 \pm 885$  &   $0.22 \sin \gamma $ &  0.28  $\pm$ 0.33   \\
\hline
\end{tabular}
\end{table}

Previous study by \citet{Pellegrini_2007} found the magnetic field strength around the southwestern part of the M17 photodissociation region with the peak value of $\sim$ 600 $ \rm \mu G$. \citet{Brogan_1999} measured directly the magnitude of the B-fields along the line-of-sight (LOS) using the Zeeman observations of \ion{H}{1} absorption lines toward the \ion{H}{2} region and obtained $ B_{\rm LOS} \sim -$450-550 $ \rm \mu G$. \citet{Chen_2012} carried out a rough estimation of the magnetic field strength based on the polarization of point sources in NIR and FIR using a sampling rectangle in the M17-S region, and they found the total magnetic field strength of $B = \sqrt{B_{\rm LOS}^2 + B_{\rm POS}^2} \sim 230 \ \mu G$ and the inclination angle of the magnetic field vector with respect to the plane-of-sky of $\sim 40^\circ$. Therefore, our estimated values of $B_{\rm POS}=980 \pm 230$ $\rm \mu G$ and $1665\pm 885$ $\rm \mu G$ with rather large uncertainties are still comparable with previous results.

Several factors affect the estimation of the magnetic field strengths. One is that we assumed spherical geometries of the molecular clumps to estimate the radii of the observed regions when estimating the volume densities. Depending on the real 3D geometry of the clouds, this assumption could lead to a large uncertainty in $B_{\rm POS}$ values. In addition, polarization observations integrate over all of the structure along the line of sight, which leads to the reduction of $\sigma_\theta$, therefore, an overestimate of B-field strengths. Also it is important to note that, statistically, margin of overestimation by the DCF method on the magnetic field strengths can be up to a factor of 2 for an individual cloud \citep{crutcher_2004}. There are efforts to improve the DCF method such as of \citet{Cho_2016} who modified the method to reduce the overestimate of $B_{\rm POS}$ strength by a factor equal to the ratio of the average line-of-sight velocity dispersion and the standard deviation of centroid velocities. We calculated the standard deviation of centroid velocities for the M17-N and M17-S which are equal to 0.59 $\rm km\; s^{-1}$ and 1.46 $\rm km\; s^{-1}$, respectively. This means that the conventional DCF method overestimates the strength of the mean magnetic field in the plane-of-the-sky by a factor of (1.6/1.46)=1.1 and (2.3/0.59)=3.9 in M17-N and M17-S, respectively (the line of sight velocity dispersion is taken from Table~\ref{tab1:summary}). Therefore, the strengths of B-fields now are $891 \pm 209 \ \rm \mu G$ and $427 \pm 227 \ \rm \mu G$ for M17-N and M17-S, respectively.

The calculated B-field strengths have quite large uncertainties which are mostly propagated from the uncertainties on column densities averaged over large regions (see the results in Table~\ref{tab1:summary}). We, therefore, examine the B-field strength limiting to the highest density regions with $N(\rm H_2) > 10^{22}\; cm^{-2}$. With this new cut, only the densest regions of M17-S are retained (see Figure \ref{fig:a4-NH2}) and the new results are $N(\rm H_2) = (7.6\pm 6.5) \times 10^{22} \rm \;cm^{-2}$, $n(\rm H_2) = (6.9 \pm 5.9) \times 10^{4} \rm \;cm^{-3}$, while \mbox{$\Delta V = (2.3 \pm 0.1) \rm \;km\; s^{-1}$} and $\rm \sigma_\theta = 6.5^\circ \pm 0.5^\circ$ are the same as before the cut is applied. Therefore, the magnetic field strength obtained is $B_{\rm POS} = 2134 \pm 738 \rm \;\mu G$ and $\lambda =0.27 \pm 0.25$ with smaller uncertainties as expected. This calculation serves as a reference providing a more precise measurement of the average magnetic strength in the dense region of M17 and a sense of possible changes of its value depending on the choices of the regions of interest. In reality, the magnetic field strength is expected to change with the local gas density.

\subsection{Alfv\'enic Mach number $\mathcal {M_A}$}
The Alfvénic Mach number, $\mathcal{M_A}$, represents the relative contribution of turbulence to magnetic fields. $\mathcal{M_A}$ is an important parameter for describing the evolution of molecular clouds \citep{Kritsuk_2017}. A sub-Alfv\'enic Mach ($ \mathcal{M_A} < 1$) means the cloud has a strong magnetic field while a super-Alfv\'enic ($\mathcal{M_A} > 1$) implies a weak magnetic field compared to turbulence. In the super-Alfvénic case, the magnetic field morphology is significantly affected by turbulence due to it subdominance. In this scenario, the morphology of the magnetic field is therefore expected to be random.

$\mathcal{M_A}$ can be calculated from the velocity dispersion following \citet{Padoan_2001, Nakamura_2008, Wang_2019} as
\begin{equation}
\label{eq:mach1}
     \mathcal{M_A} = \frac{\sigma_\nu}{ v_A} = \frac{ \sigma_\nu  \sqrt{ 4 \pi \rho }}{ B },
\end{equation}
where $v_A$ is the Alfv\'enic velocity. Combining with Equation \ref{eq:B_mag}, we obtain
\begin{equation}
\label{eq:mach2}
     \mathcal{M_A} = \frac{\sigma_\theta \sin \gamma}{Q_c},
\end{equation}
where $\gamma$ is the inclination angle of B-fields with respect to the line-of-sight in the range of $ [0^\circ,90^\circ]$ with $B_{\rm POS} = B \sin \gamma$, and $Q_c=0.5$ \citep{Ostriker_2001}. 

Using the inferred polarization angular dispersion in Table \ref{tab1:summary}, we obtained $\mathcal{M_A} = 0.12 \sin \gamma$ and $0.22 \sin \gamma$ for M17-N and M17-S, respectively. Thus, M17 is sub-Alfv\'enic, implying that the magnetic fields in the region dominate turbulence. The sub-Alfv\'enic Mach numbers are also in agreement with the generally well-ordered magnetic field morphology in the region (see Figure \ref{fig8:M17_halfvector}).

\subsection{Mass-to-flux ratio}
The mass-to-flux ratio, $M/ \Phi$, refers the ratio of the mass to the magnetic flux. This is a crucial parameter describing the role of B-fields relative to gravity in star formation \citep{crutcher_richard}. Usually, it is determined by the critical value, defined as follows \citep{crutcher_2004}
\begin{equation}
\label{eq:lambda}
\lambda = \frac{(M/\Phi)_{\rm observed}}{(M/\Phi)_{\rm critical}} = 7.6 \times 10^{-21} \frac{ \langle N(\rm H_2) \rangle }{B_{\rm POS} },
\end{equation}
where the critical mass-to-flux ratio $ (M/\Phi)_{\rm critical} = \dfrac{1}{2 \pi \sqrt{G} } $ \citep{Nakano_1978}, $\langle N(\rm H_2) \rangle$ is the average column density of the considered region in units of $ \rm cm^{-2}$, and $B_{\rm POS}$ is given in units of $ \rm \mu G$.\footnote{We note that an analysis by \cite{crutcher_2004} indicated that statistically the true mass-to-flux ratio can be over-estimated by a factor of three.} The errors on the mass-to-flux ratio are calculated using the following equation
\begin{equation}
\label{eq:error}
    \frac{\sigma_\lambda}{\lambda} =  \sqrt{  \left(   \dfrac{\sigma_{N(\rm H_2)}}{ \langle N(\rm H_2) \rangle} \right)^2  + \left(  \frac{\delta B_{\rm POS}}{ B_{\rm POS}}  \right)^2   }.
\end{equation}

A super-critical value of $\lambda > 1$ means that the gravity dominates the magnetic field pressure, and the cloud can undergo gravitational collapse to form a protostar. Conversely, a sub-critical value of $ \lambda < 1$ indicates that the magnetic field is strong enough to counteract the gravitational collapse \citep{crutcher_2004, Pattle_2017}. 

We obtained $\lambda= 0.07$ and 0.28 for the M17-N and M17-S regions, respectively (see Table \ref{tab:Bpos}). Overall, the M17 cloud is sub-critical. It means that M17 is magnetically supported and belongs to the strong magnetic field model of star formation theory \citep{Nakano_1978}. However, the $\lambda$ values here are calculated by averaging over the whole considered regions. Therefore, the highest density cores can still become super-critical and then gravitationally collapse to form new stars. The inferred values of the mass-to-flux ratio are compatible with theoretical predictions of the am-bipolar diffusion model where the B-field morphology is dragged inward from the outer layers of the cloud to the massive cores in both M17-N and M17-S. In addition, these results are compatible with the deficiency of forming massive stars and the lack of gravitationally bound clumps in the regions \citep{Nguyen_Luong_2020}.

\begin{figure*}[htbp!]
\centering
\includegraphics[width=1.0\textwidth]{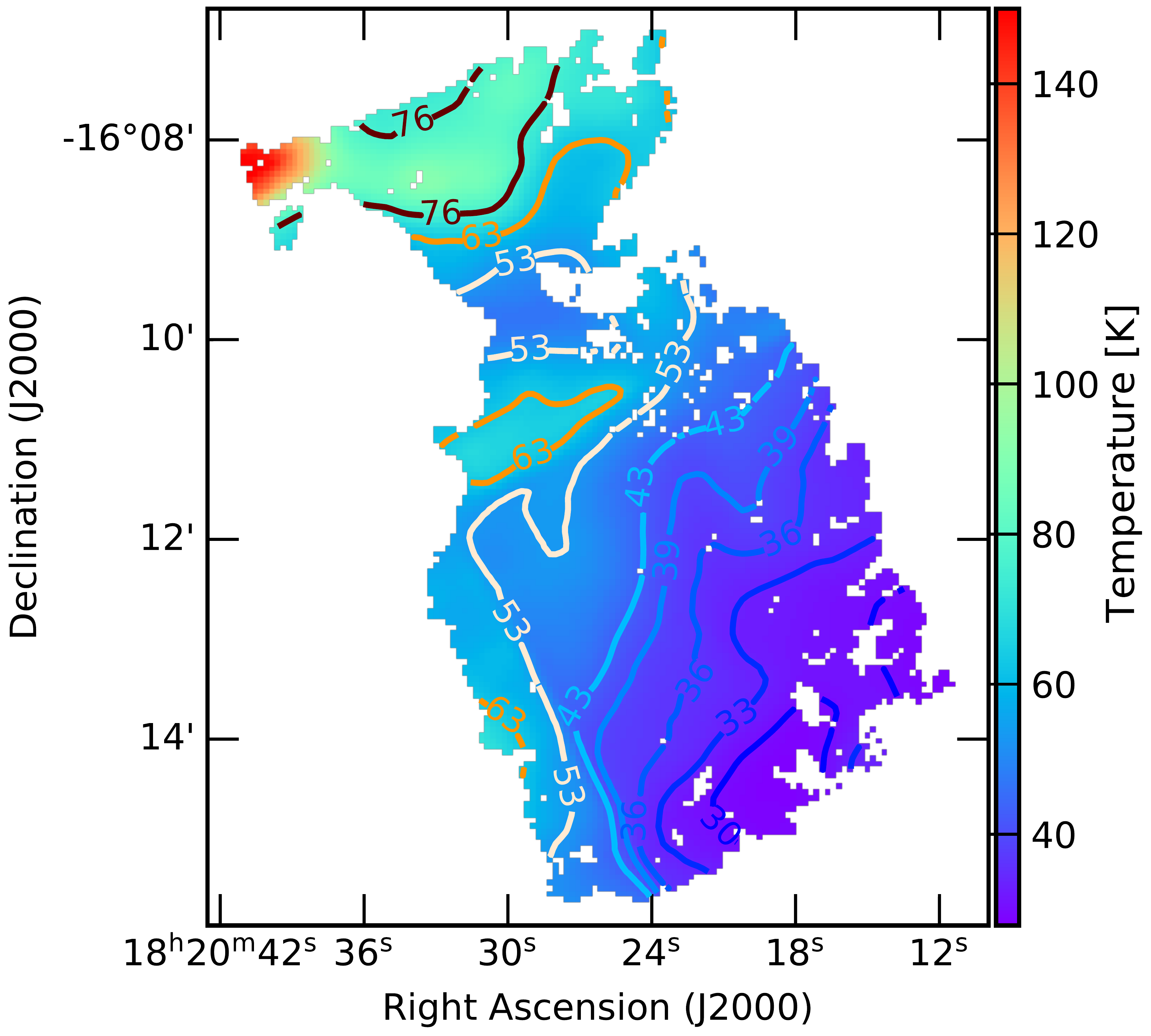}
\caption{Dust temperature map of M17 with the contours indicating the corresponding temperatures. In M17-S, the dust temperature tends to decrease from the east to the west, in the direction away from the star cluster.}
\label{fig10:M17_temp}
\end{figure*}

\subsection{Stellar feedback from the massive star cluster}\label{sec:feedback}
As shown in Figures \ref{fig8:M17_halfvector} and \ref{fig9:M17_steplic}, the magnetic fields in the vicinity of the star cluster (yellow symbols in Figure \ref{fig9:M17_steplic}) appear to be bent and aligned with the density structure (see more in Section \ref{subsec:bfieldmorph}). This suggests the effects of stellar feedback of the massive star cluster.

To understand the importance of feedback, we estimate the ratio of the ram pressure, $P_{\rm ram}$, by stellar winds (or expansion of HII region) to the magnetic pressure, $P_{\rm mag}$, which is given by
\begin{eqnarray}
\label{eq:Pratio}
\frac{P_{\rm ram}}{P_{\rm mag}}&=&\frac{\rho v_{\rm wd}^{2}}{B^{2}/8\pi} \nonumber\\
&\simeq& 4.7\left(\frac{v_{\rm wd}}{20~\rm km\;\rm s^{-1}}\right)^{2} \left(\frac{1000~\mu {\rm G}}{B}\right)^{2}\left(\frac{n({\rm H}_{2})}{10^{4}~\rm cm^{-3}}\right),~~~~~~~
\end{eqnarray}
where the wind speed $v_{\rm wd}\sim 20\rm \;km~s^{-1}$ is adopted from \cite{Pellegrini_2007}. The mass density $\rho=2.8m_{\rm H}n({\rm H}_{2})$. The volume densities, $n({\rm H}_{2})$, and magnetic field strengths, $B$, close to those from Table \ref{tab1:summary} and \ref{tab:Bpos} have been used. 
The dominance of the ram pressure over the magnetic pressure as shown in Equation \ref{eq:Pratio} implies that the winds can bend the magnetic fields frozen in the gas. Stellar feedback due to the star cluster may also trigger star formation in the M17-S region, as revealed by the presence of numerous IR sources. 

\subsection{Dust temperature map}

As described in Section \ref{sec:nh}, the dust temperature, $T_{\rm d}$, map has been derived via the graybody fit with {\it Herschel} 160-$500\;\micron$ data. Then, the template $T{\rm _d}$-map has been re-gridded to match to SCUBA2 $850\;\micron$ map where the final map has the pixel size of 4$\arcsec$.

Figure \ref{fig10:M17_temp} shows the temperature map of the observed region by {\it Herschel}. The average temperatures are $74 \pm 11$ and $43 \pm 10$ K for the M17-N, and M17-S, respectively. Globally, the M17 dust temperature gradually decreases from east to west and from north to south. This general trend is consistent with the scenario that dust is heated by the massive star cluster (see e.g., Figure \ref{fig9:M17_steplic}). The higher temperatures in M17-N and the eastern and northeastern parts of M17-S arise from the fact that these regions are located next to the star cluster and thus strongly heated by this intense radiation source as well as the \ion{H}{2} region in the east. M17-N is additionally affected by shocks from G015.128 resulting in an obvious temperature excess. In the M17-S, the dust temperatures decrease along the direction away from the star cluster, which stems from dust absorption within dense structures of an order of magnitude higher than that of M17-N.

\begin{figure*}[htbp!]
\centering
\includegraphics[width=0.9\textwidth]{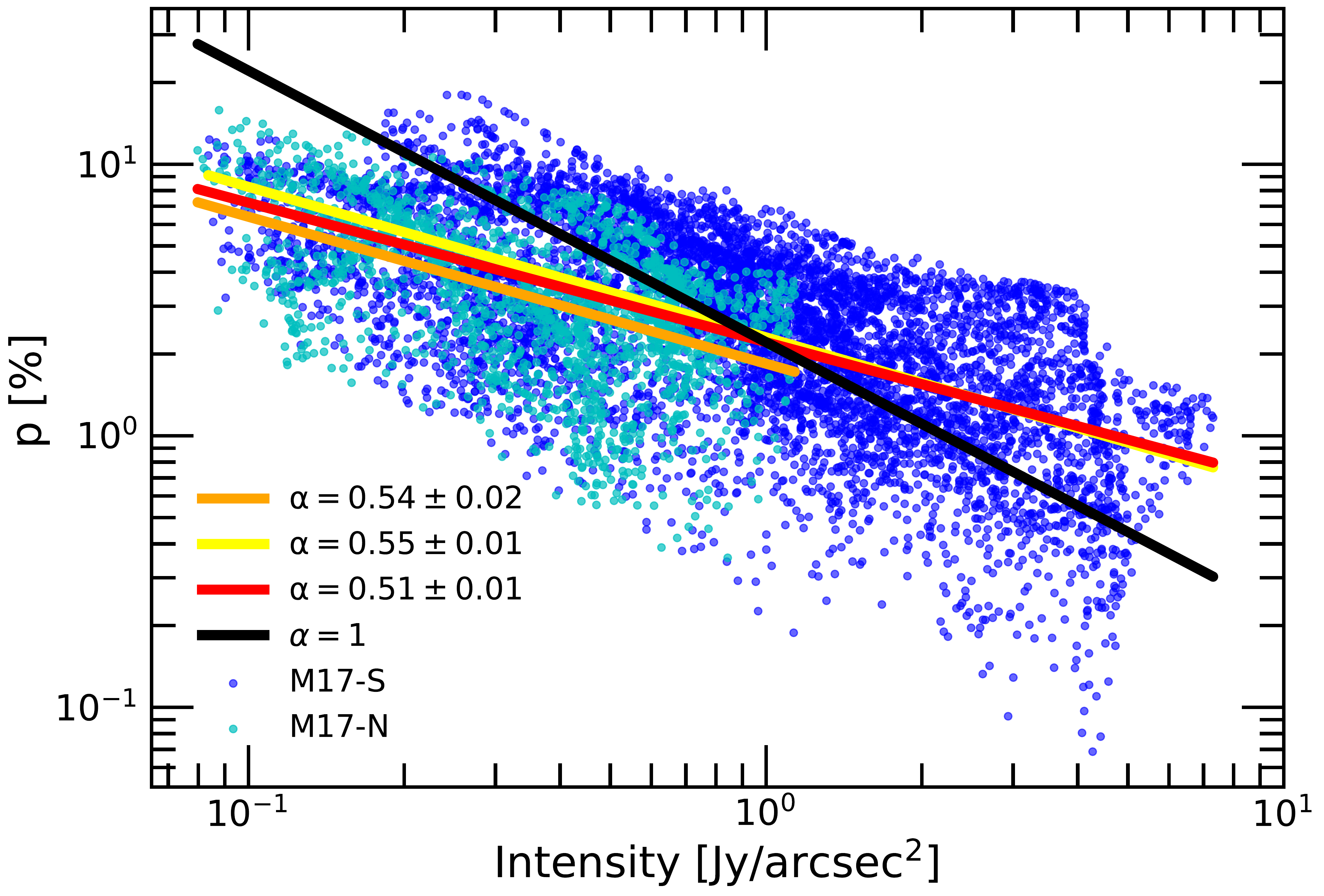}
%\plotone{SOFIA_bandD_M17_I_p_fit_236_legend_title_regions.pdf}
\caption{The variation of the polarization fraction with the total intensity. The orange, yellow and red lines are the best fit to a power law model for M17-N, M17-S and the whole map, respectively. The black solid line is the same power law model with $\alpha = 1$.}
\label{fig:M17_I_p_fit}
\end{figure*}

\subsection{Polarization fraction vs. emission intensity, column density, and dust temperature}
\label{sub:grains}
We now study the dependence of the polarization fraction, $p$, on the total intensity, $I$, the column density, $N({\rm H_2})$, and the dust temperature, $T_{\rm d}$, which reveal basic properties of grain alignment and disruption in M17.

\begin{figure*}[htbp!]
\centering
    \includegraphics[width=0.49\textwidth]{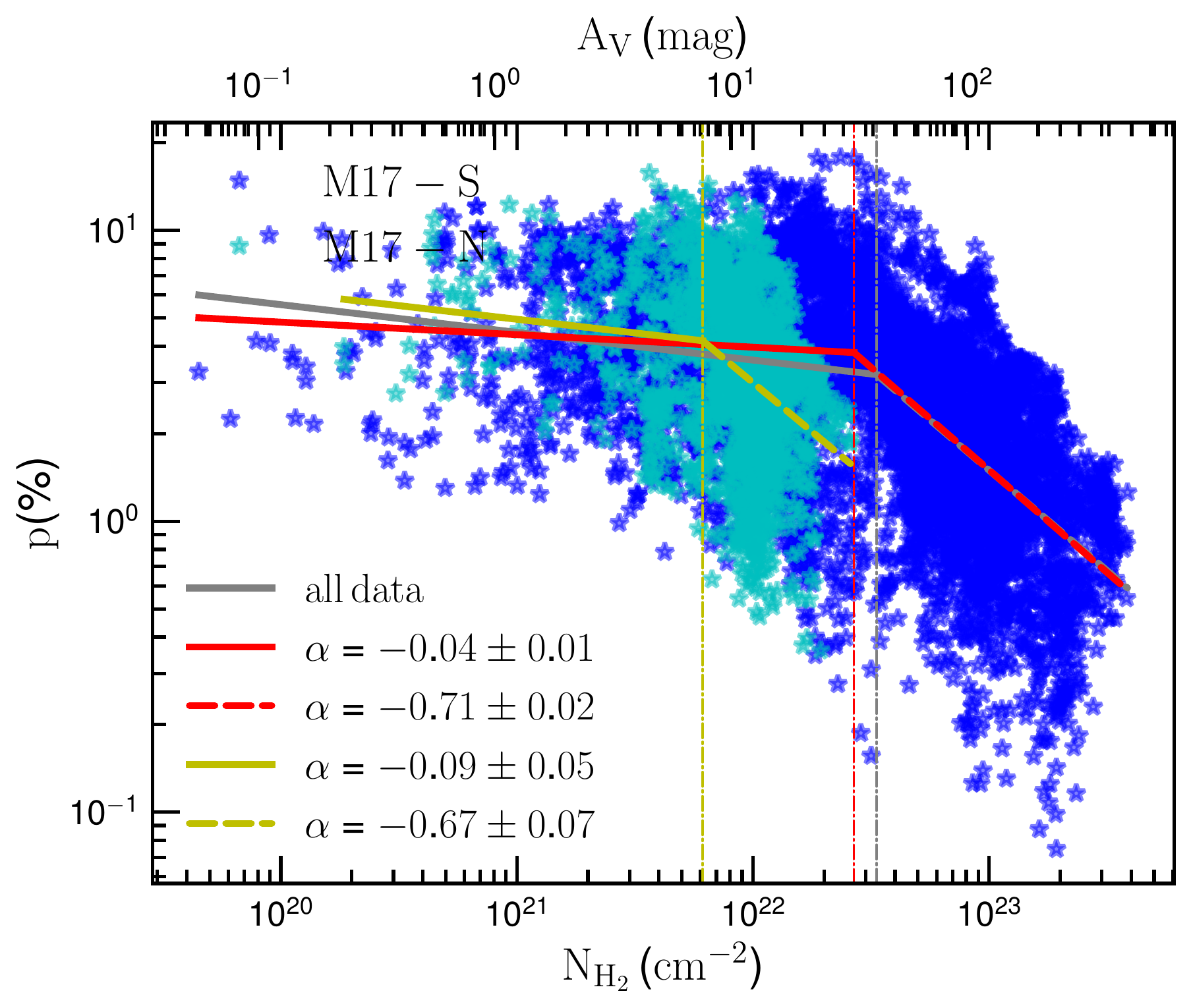}
    \includegraphics[width=0.49\textwidth]{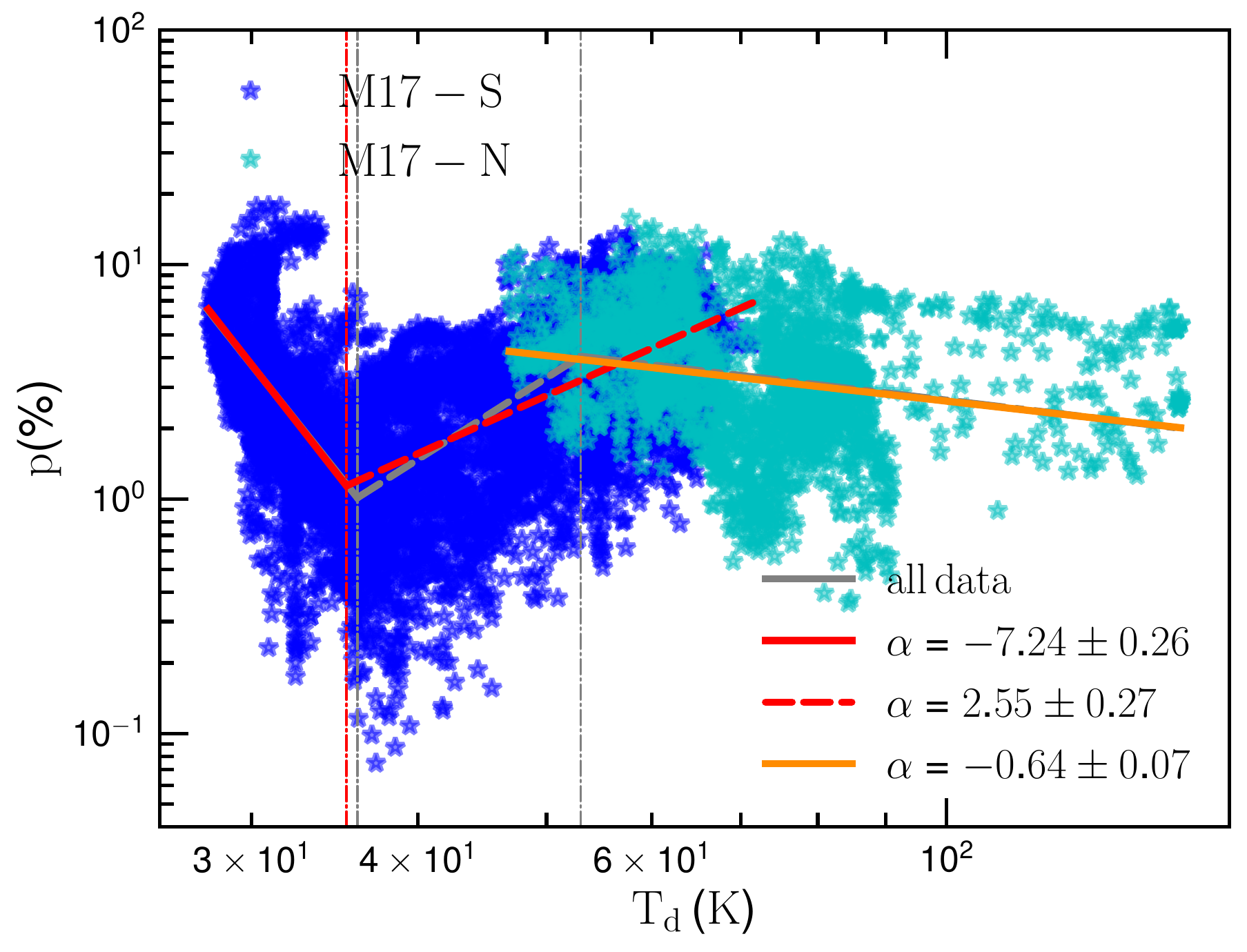}
    \includegraphics[width=1.0 \textwidth]{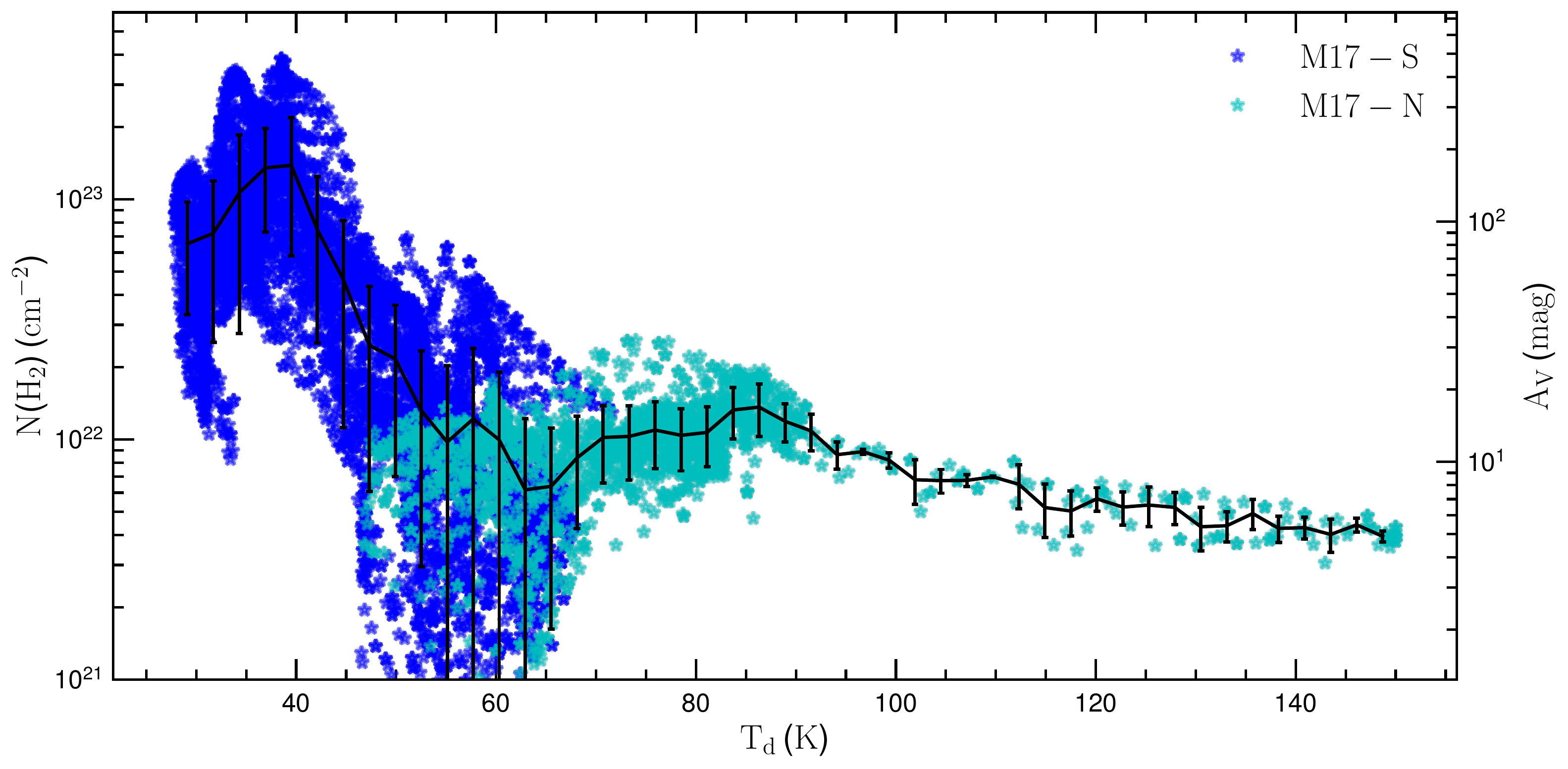}
\caption{Upper-left: the relationship of the polarization fraction and column density. Upper-right: the relationship of the polarization fraction and temperature. $\alpha$ is the power index of a power-law model. The SOFIA/HAWC+ polarization map is smoothed to 14\arcsec angular resolution of SCUBA2 $850\;\micron$ which were used together with Herschel data to get the $N({\rm H_2})$ and $T_{\rm d}$ maps. Bottom: the variation of the gas column density, $N({\rm H_2})$, with the dust temperature, $T_{\rm d}$.}
\label{fig:p_T_NH2}
\end{figure*}

\begin{figure*}[htbp!]
\centering
\includegraphics[trim=0.7cm 0.2cm 1.3cm 0.7cm,clip,width=0.49\textwidth]{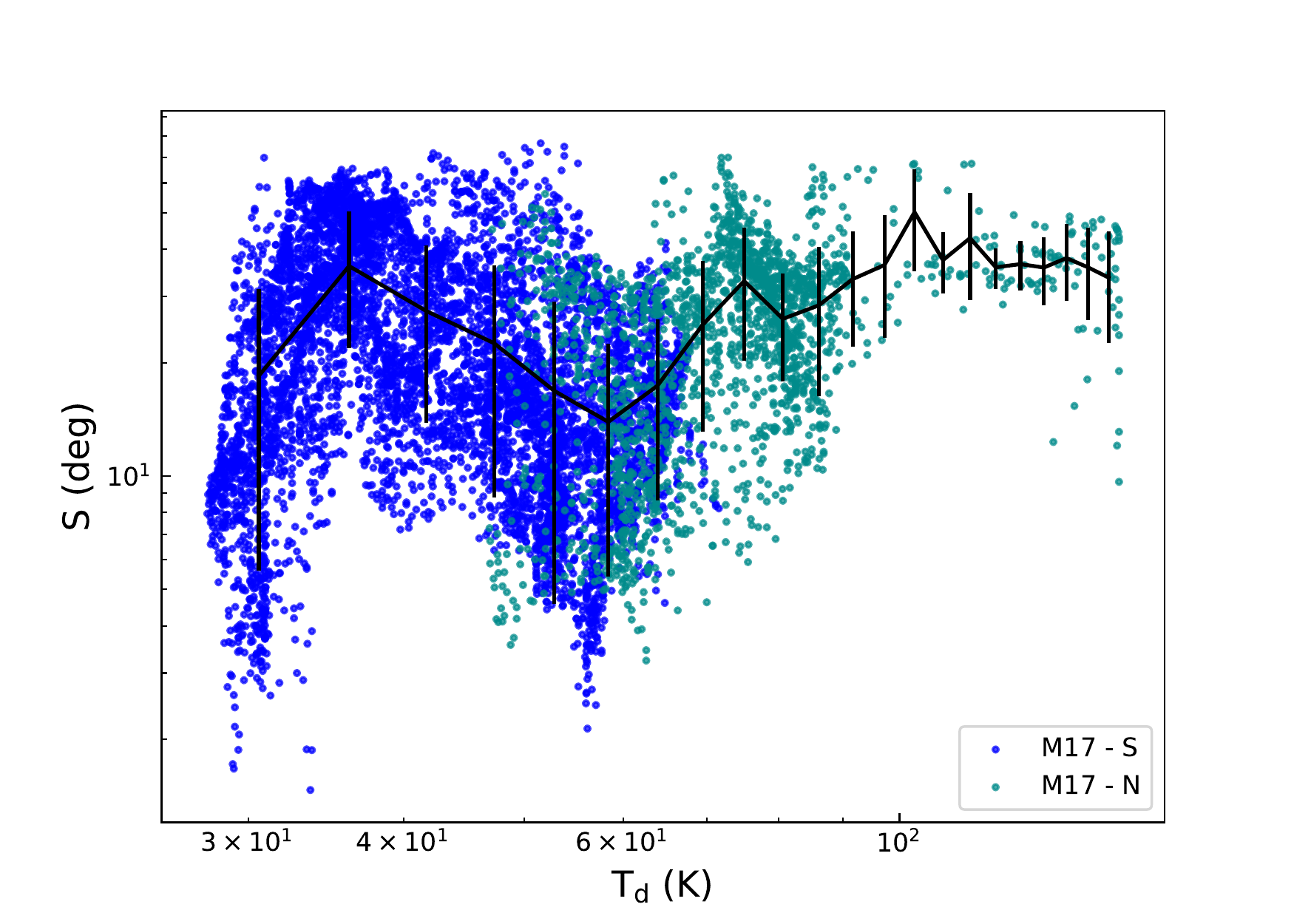}
\includegraphics[trim=0.7cm 0.2cm 1.3cm 0.7cm,clip,width=0.49\textwidth]{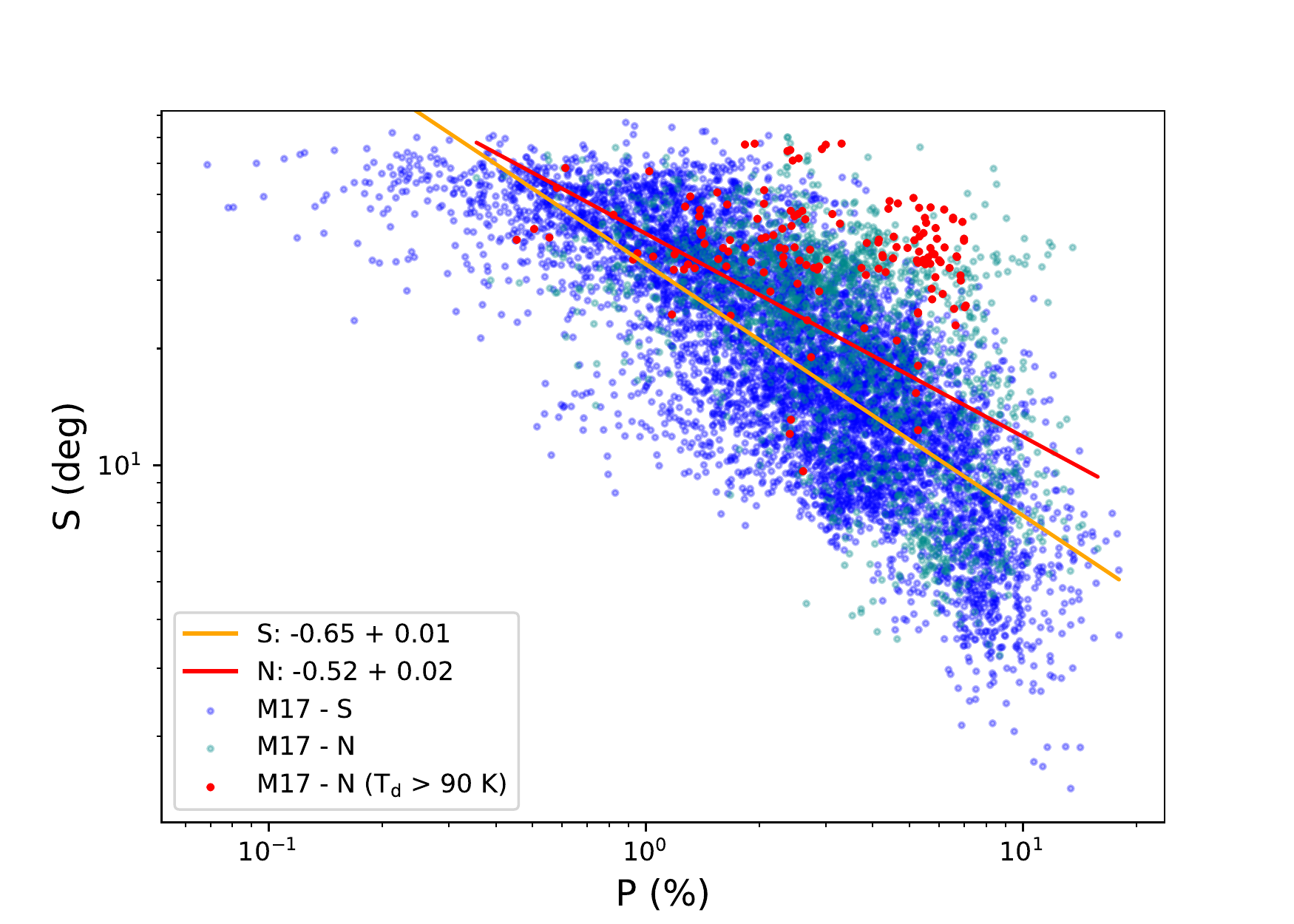} 
\caption{Left: Variation of the polarization angle dispersion function, $S$, with the dust temperature, $T_{\rm d}$. The black
 curve shows the mean and the standard deviation per bin. Right: Variation of $S$ with the polarization fraction, $p$. Red dots are pixels with \mbox{$T_{\rm d} > ~90$ K}. The orange and red lines are the fit to a power law model for M17-N and M17-S, respectively.} \label{fig:svst}
\end{figure*}

Firstly, we investigate the relationship between the polarization fraction, $p$, and the total intensity, $I$. It is clearly seen from Figure \ref{fig3:M17_pol_fulmap} that polarization fraction decreases when intensity increases. Generally, the dependence of $p\propto I^{-\alpha}$ describes the variation of the grain alignment efficiency and the magnetic field geometry across the cloud. For a uniform magnetic field, the slope $\alpha=1$ implies that the grain alignment is present only in the outer layer of the cloud and becomes completely lost in the inner region. Figure \ref{fig:M17_I_p_fit} shows a fitted power-law model $p\propto I^{-\alpha}$ with the power index $\alpha = 0.51 \pm 0.01$ for the whole M17 region. The values of $\alpha$ do not change much when we fit only with pixels from M17-N ($\alpha=0.54 \pm 0.02$) or M17-S ($\alpha = 0.55 \pm 0.01$) separately. A best-fitted value of $\alpha=0.51$ reveals that grain alignment is significant toward the high emission intensity. As a conclusion, we observe that the polarization fraction, $p$, decreases with increasing the emission intensity, $I$. 

To better understand the variation of $p$, we now study the relationship between the polarization fraction, $p$, and the column density, $N({\rm H_2})$, as well as the dust temperature, $T_{\rm { d} }$. Figure \ref{fig:p_T_NH2} (top-left) shows the $p-N(\rm { H_{2} })$ relation. Here we derive the slopes by fitting piece-wise linear functions to the data. In the M17-S region, the polarization fraction gradually decreases with increasing the gas column density, before it experiences a steep drop with a slope of $-0.71$ for $N_{\rm H_{2}}>3\times 10^{22}\rm \; cm^{-2}$ or the visual extinction\footnote{ $A_{\rm V}(\rm mag)=R_{V}\times 2N({\rm H_2})/(5\times 10^{21}\;{\rm cm^{-2}}) $} $A_{\rm V} >37$ mag for the typical total-to-selective extinction $R_{V}\approx 3.1$.
In the M17-N region, a steep decrease with slope of $-0.67$ occurs at $N({\rm H_2}) > 5 \times 10^{21}\;\rm cm^{-2}$ or $A_{\rm V}>6$ mag. The strong depolarization occurs in the region where the gas density is relatively low and the dust temperature is relatively high compared to that of M17-S (see Figure \ref{fig10:M17_temp}), which is unexpected from the RATA theory.    

The top-right panel of Figure \ref{fig:p_T_NH2} shows the $p-T_{\rm d}$ relation. It appears that the polarization fraction first decreases, then increases before decreasing again as the dust temperature increases. The decrease-increase feature originates from the M17-S region, while the depolarization is from M17-N. 

The bottom panel of Figure \ref{fig:p_T_NH2} shows the gas column density as a function of the dust temperature. The column density decreases rapidly with increasing $T_{d}$ in M17-S, but it varies slowly in M17-N. The first decrease of $p$ for $T_{\rm d} < 40\,\rm K$ corresponds to the highest gas density ($A_{\rm V} \sim 100$ mag). This depolarization is caused by decrease of grain alignment due to the attenuation of the radiation field and the enhancement of collisional damping. Towards higher dust temperatures, the gas density drops, and the grain alignment efficiency enhances toward the central luminous source, which results in the following increment of $p$. These features are expected from the context of the RATA theory. In M17-N, the dust temperature could be up to 150$\,$K and the gas density becomes more diffuse at $A_{\rm V}<10$ mag, the polarization fraction, however, appears to monotonically decrease to higher $T_{\rm d}$, which is completely contradictory to what expected from RATA theory. 

\subsection{Implications for grain alignment and rotational disruption by RATs}

We now discuss the implications of the observed polarization fraction toward M17 for physics of grain alignment and disruption based on radiative torques. 

According to the RATA theory (see \citealt{Lazarian2007}), the polarization fraction of thermal emission increases with decreasing alignment size, $a_{\rm align}$, which is the minimum size of aligned grains. The alignment size is determined by the balance between spin-up by RATs and spin-down by gas collisional damping, which is a function of the local dust temperature (or radiation intensity) and gas density as $a_{\rm align} \sim n^{2/7}_{\rm H}T^{-12/7}_{\rm d}$ (\citealt{tram2021sofia}). As a result, a denser gas and lower dust temperature increase $a_{\rm align}$ (see details in \citealt{Hoang_2021}), which results in the decrease of the polarization fraction. Such a prediction by the RATA theory can explain the decrease of the polarization fraction with increasing the gas column density observed in the \mbox{M17-S} region (see Figure \ref{fig:p_T_NH2}; left panel).
However, it can not explain the decrease of $p$ with increasing $T_{\rm d}$ in the M17-N region where the gas density changes slowly (see Figure \ref{fig:p_T_NH2}; right panel), which reveals evidence of the RATD.

The polarization fraction at far-IR/submm is very sensitive to the maximum size of the grain size distribution because large grains dominate long-wavelength emission. Following the RATD mechanism (\citealt{Hoang_2019}), the maximum grain size above which large grains are disrupted by RATs is determined by the local dust temperature (radiation field), gas density, and the grain tensile strength, $S_{\rm max}$, which follows by $a_{\rm disr} \sim n^{1/2}_{\rm H}T^{-3}_{\rm d}S^{1/4}_{\rm max}$ (\citealt{tram2021sofia}). For the average estimated volume density $n_{\rm H_{2}}\simeq 3\times 10^{4}\,\rm cm^{-3}$ (see Table \ref{tab1:summary}), and the temperature at the breaking point in Figure \ref{fig:p_T_NH2} of $T_{\rm d}\simeq 53\,\rm K$, one obtains $a_{\rm disr}=0.16$, $0.28$, and $0.49\,\mu$m for $S_{\rm max}=10^{7}$, $10^{8}$, and $10^{9}\,\rm erg\,cm^{-3}$, assuming the mean wavelength of radiation field of $1\,\mu$m. The decrease of the disruption size by RATD leads to the decrease of the dust polarization as the dust temperature increases (\citealt{Lee_2020}; \citealt{Tram2021b}), which successfully reproduces the observed trend toward the M17-N (see Figure \ref{fig:p_T_NH2}).

To quantify the magnetic field tangling at small scales on the depolarization, we calculate the polarization angle dispersion function, $S$, (see Section 3.3 in \citealt{planck2015_dust_pol}). For each pixel at location $x$, $S$ is calculated as the standard deviation of the polarization angle difference, $S_{xi}$, of pixel $x$ and pixel $i$ which lies on a circle having $x$ as the center and a radius of $\delta$
\begin{equation}
\label{eq:ang_dis_func}
 S^2(x,\delta) = \frac{1}{N}
\sum _{i=1}^{N}S_{xi}^2,
\end{equation}
where $S_{xi}=\theta(x)-\theta(x+\delta)$ is the polarization angle difference and $N$ is the number of pixels lying on the circle.

For the current SOFIA/HAWC+ data set, we calculate $S$ for $\delta = 27.2\arcsec$ ($\sim$ 2 beam sizes). The left panel of Figure \ref{fig:svst} shows the $S$ - $T_{\rm d}$ relation where it can be seen that the angular dispersion function $S$ strongly correlates with the dust temperatures but not the highest ones, \mbox{$T_{\rm d} > ~90$ K}, in M17-N. Thus, the decline of $p$ at \mbox{$T_{\rm d} > ~90$ K} (see the upper-right panel of Figure~\ref{fig:p_T_NH2}) cannot be explained by the magnetic field tangling and presents an evidence for the RATD effect. The right panel of Figure \ref{fig:svst} shows a general anti-correlation of the angular dispersion function $S$ and polarization fraction $p$, except, again, for the highest temperatures pixels (red dots) whose temperatures are greater than 90 K. For \mbox{$T_{\rm d} < ~35$ K} located in M17-S (blue dots in Figure~\ref{fig:p_T_NH2} and ~\ref{fig:svst}), we can see that the rapid decrease of $p$ is due to B-field tangling in the high density region (see bottom panel of Figure~\ref{fig:p_T_NH2}). For the temperature range from \mbox{35 K} to 65 K, in both M17-N and M17-S, $N({\rm H_2})$ and $S$ decrease, therefore, $p$ increases. This clearly shows the important contribution of field tangling to the depolarization at low dust temperatures.

A detailed modeling of dust polarization using the grain alignment and disruption by RATs for M17 is beyond the scope of this paper, and will be addressed in a follow-up study.  

\section{Summary} 
\label{subsec:summary}
In this study, we use the dust polarization data taken by SOFIA/HAWC+ at $154 \ \rm \mu m$ to study magnetic fields and dust physics in the M17 nebula. 
Our main results are summarized as follows:
\begin{enumerate}
    \item 
Using the DCF method, we estimated the magnetic field strengths to be $980 \pm 230 \ \rm \mu G$ in the lower-density region (M17-N) and $1665 \pm 885 \ \rm \mu G$ in the higher-density region (M17-S). The strong magnetic field could be a result of the pressure exerted by the \ion{H}{2} region in the eastern part of the observed region by SOFIA/HAWC+ toward M17. In the M17-N, the B-field morphology can be that of a gravitational collapse molecular core. The morphology of B-fields in M17-S displays a well-organized elongated structure along with the gravitational collapse directions from the outer regions to the denser central regions. The fields are dragged inward to the gravitational center. The whole B-field morphology represents an asymmetric large-scale hourglass structure. We also found a pillar structure which is one of the common features of PDR regions. A rough estimation of the contribution of ram over the magnetic pressure suggests that the wind from the PDR region is strong enough to impact on the B-field morphology of M17.

\item
The Mach numbers are determined to be sub-Alfv\'enic ($\mathcal{M_A} < 1$) which indicate that the magnetic field dominates turbulence. In addition, the inferred sub-critical values of mass-to-flux ratios, $\lambda < 1$, imply that the magnetic fields in the regions are strong enough to resist gravitational collapse. These results are consistent with the deficiency of the formation of massive stars in the region from previous studies.

\item
To study dust physics, we analyzed the relation of the polarization fraction, $p$, with the emission intensity, $I$, gas column density, $N(\rm H2)$, and dust temperature, $T_{d}$. The power index of the $p$ vs $I$ relation, $\alpha = 0.51$, implies that dust grains could still be aligned by radiation in the region. The decrease of the dust polarization with the column density in the M17-S can be explained by the RATA theory as well as the tangling of the magnetic field. 

\item To study the effect of magnetic field tangling on the dust polarization, we also analyzed the variation of the polarization angle dispersion function, $S$, with dust temperature and gas column density. In the M17-N, the decrease of $p$ with $T_{d}$ at high temperatures when both $N(H_{2})$ and $S$ decrease is most consistent with the theoretical predictions of dust polarization by both RATA and RATD effects.

\item
There are large statistical biases on the estimation of the polarization angle dispersion, $\sigma_\theta$, as well as the volume density, $n(\rm H_2)$, due to the fact that the magnetic field strengths are estimated over a large areas. These biases lead to large uncertainties on the measurement of the magnetic field strengths. It is also known that for a random sample of magnetic field orientations, using the DCF method the mass-to-flux ratio will on average be overestimated by a factor $\sim 3$. In fact, using a method proposed by \cite{Cho_2016} we found the overestimate factors equal to 1.1 and 3.9 for the M17-N and M17-S, respectively. Therefore, new methods of estimating magnetic field strength which are able to identify the fields and accompanied useful parameters such as Alfv\'enic Mach numbers and mass-to-flux ratios in the basis of pixel by pixel would provide detailed and more precise conditions of star formation in the M17 molecular cloud.
\end{enumerate}

\acknowledgments
This research is based on observations made with the NASA/DLR Stratospheric Observatory for Infrared Astronomy (SOFIA). SOFIA is jointly operated by the Universities Space Research Association, Inc. (USRA), under NASA contract NNA17BF53C, and the Deutsches SOFIA Institut (DSI) under DLR contract 50 OK 0901 to the University of Stuttgart. Nobeyama Radio Observatory is a branch of the National Astronomical Observatory of Japan, National Institutes of Natural Sciences, part of the data were retrieved from the JVO portal (http://jvo.nao.ac.jp/portal/) operated by ADC/NAOJ. We thank Joseph M. Michail for his line integral convolution (LIC) Python tool. T.H. acknowledges the support by the National Research Foundation of Korea (NRF) grants funded by the Korea government (MSIT) through the Mid-career Research Program (2019R1A2C1087045). P.N.D., N.B.N., and N.T.P. are grateful to the funding from the Vietnam National Foundation for Science and Technology Development (NAFOSTED) under grant number 103.99-2019.368. N.L. acknowledges the support from the First TEAM grant of the Foundation for Polish Science No. POIR.04.04.00-00-5D21/18-00. K.P. is a Royal Society University Research Fellow.

\vspace{5mm}
\facilities{SOFIA HAWC+, Nobeyama 45m Telescope, GLIMPSE Spitzer Data, Herchel}

\software{spectral\_cube \citep{spectral_cube}, aplpy \citep{aplpy2019},  astropy \citep{astropy:2013, astropy:2018}, reproject \citep{reproject_tool} }

%\newpage
%\clearpage
\bibliography{bibliography}{}
\bibliographystyle{aasjournal}

%% =============================================================
%% ===========APPENDIX============================
%% ============================================================
%\newpage
%\clearpage
\appendix
\counterwithin{figure}{section}
\counterwithin{table}{section}
\section{Appendix}
\label{appendixA}
This appendix shows the characteristics of the observational data and the behavior of the master cut applied to the data. 

\begin{figure}[htbp!]
\includegraphics[width=1.\textwidth]{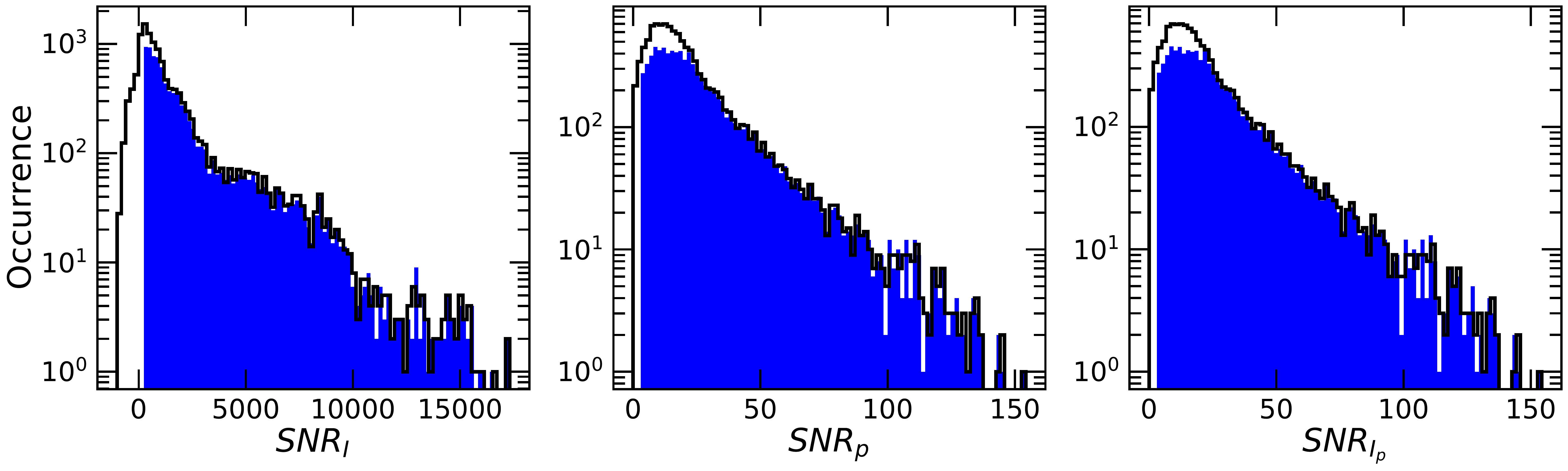}
\caption{Left to right: Distributions of signal-to-noise ratios of the total intensity ($\rm  SNR_I$), polarization fraction ($\rm SNR_p$), and polarization intensity ($\rm SNR_{I_p}$). The histograms are before cut and blue parts after cut.}
\label{fig:M17_SNR_cut}
\end{figure}

Figure \ref{fig:M17_SNR_cut} shows the distributions of the SNRs of $I$, $p$, and $I_p$ before and after applying the master cut defined in the text. Their corresponding mean and RMS values are listed in Table \ref{tab:SNR}. The mean values of the SNRs increase significantly after the cut by 45\%, 18\%, and 17\% for $I$, $p$, and $I_p$, respectively.
\begin{table}[htbp!]
\centering
\caption{Means and RMS of the distribution of $\rm SNR_I, \ SNR_p, \ SNR_{I_p}$. }
\label{tab:SNR}
\begin{tabular}{|c|c|c|c|c|} 
\hline
 \multicolumn{2}{|c|}{} &  $\rm SNR_I$ & $\rm SNR_p$ & $\rm SNR_{I_p}$  \\ 
\hline
\multirow{2}{*}{Before cut} & Mean & 1477.4  & 22.2 &   22.3 \\
                            & RMS &  2186.3 & 19.6  &  19.6  \\
\hline
\multirow{2}{*}{After cut} & Mean &  2138.5 & 26.1  &  26.1     \\ 
                            & RMS &  2366.7 &  20.8 &    20.9  \\
\hline
\end{tabular}
\end{table}

\begin{figure}[htbp!]
\centering
\includegraphics[width=0.85\textwidth]{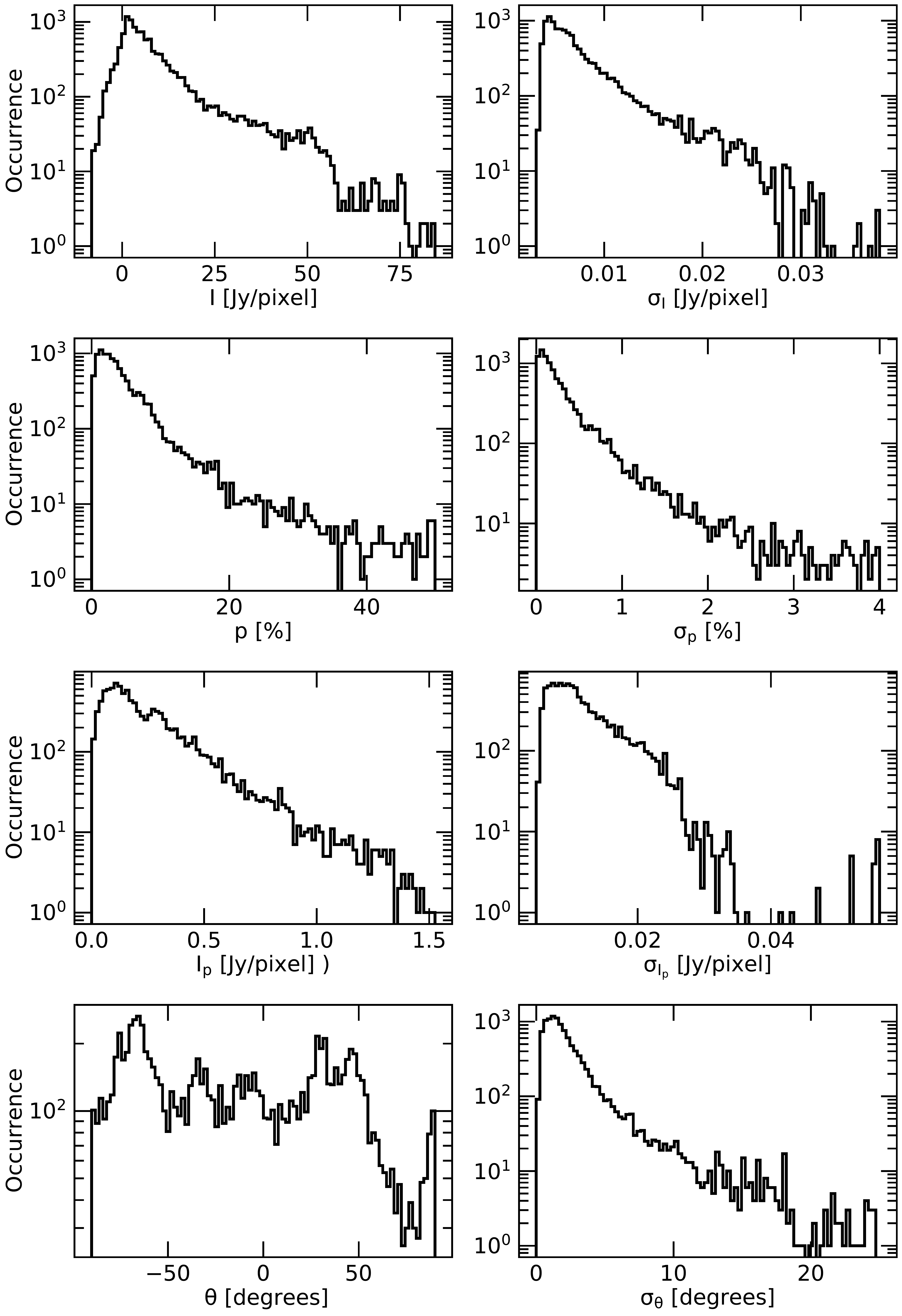}
\caption{From left to right, top to bottom: Distributions of I, $ \sigma_I$, p, $ \sigma_p$, $ I_p$, $ \sigma_{I_p}$, $ \theta$, and $ \sigma_{\theta}$ of the raw SOFIA/HAWC+ data.}
\label{fig:M17_distribute}
\end{figure}

Figure \ref{fig:M17_distribute} presents one dimensional histograms of the raw data of $ I, \ \sigma_I$, $ \ p, \ \sigma_p$, $I_p, \ \sigma_{I_p}$, $ \theta, \ \sigma_{\theta}$. The associated mean and RMS values of these quantities presented in Table \ref{tab:data}.

\begin{table}[htbp!]
%\centering
\caption{Mean and RMS values of $ I, \ \sigma_I, \ I_p, \ \sigma_{I_p}, \ p, \sigma_{p}, \ \theta$, and $\ \sigma_{\theta}$. }
\label{tab:data}
\begin{tabular}{|c|c|c|c|c|c|c|c|c|} 
\hline
\hline
& $ I$ & $ \sigma_I$ & $ I_p$ & $ \sigma_{I_p}$ & $ p$  & $ \sigma_p$ & $ \theta$ & $ \sigma_{\theta}$  \\ 
& $\rm [Jy/pixel]$ & $\rm [Jy/pixel]$ & $\rm [Jy/pixel]$ & $\rm [Jy/pixel]$ & $ [\%]$  & $ [\%]$ & $ [^\circ]$ & $ [^\circ]$  \\ 
\hline
Mean & 9.4 & 0.0078 & 0.247 & 0.0119 & 5.0 & 0.367 & -11.1 &  2.5 \\ 
\hline
RMS & 12.7 & 0.0046 & 0.215 & 0.0054 & 5.9 & 0.507 & 48.9 & 2.8 \\ 
\hline
\end{tabular}
\end{table}

\begin{figure*}[htbp!]
\centering
    \includegraphics[width=0.6\textwidth]{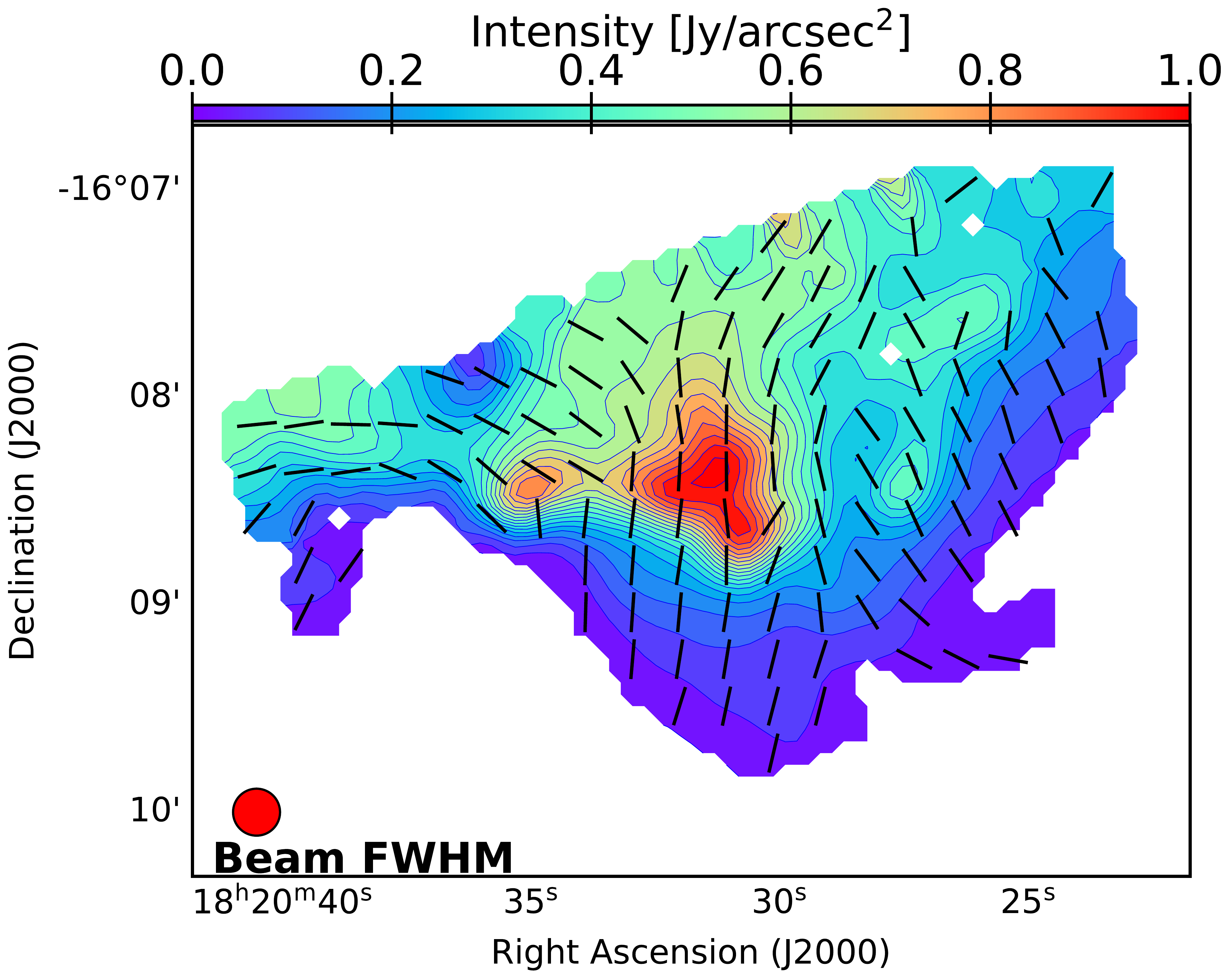}
    \includegraphics[width=0.6\textwidth]{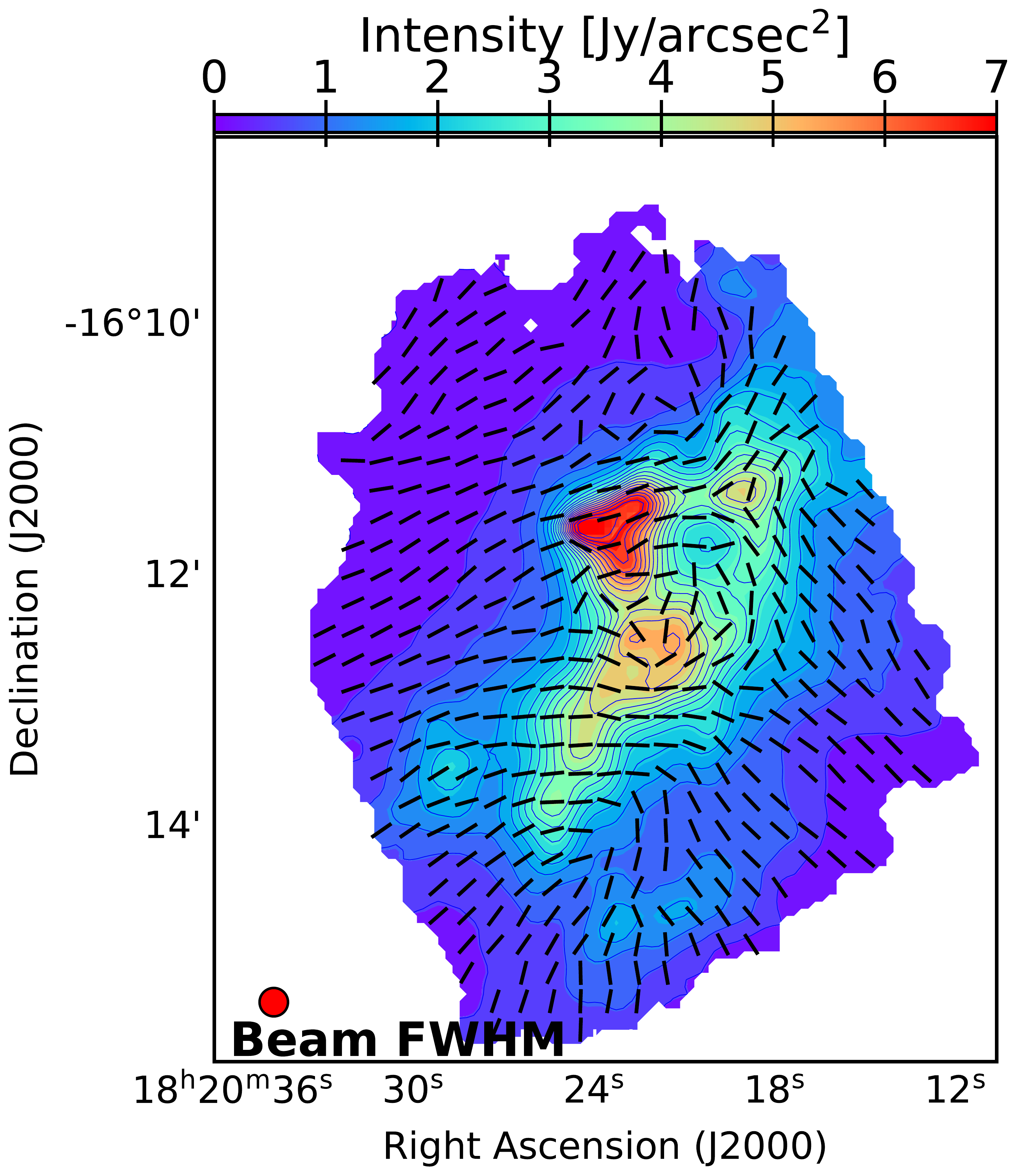}
\caption{A close-up view of B-fields of the SOFIA/HAWC+ M17 regions: M17-N (top) and M17-S (bottom). Description of the maps is the same for Figure \ref{fig3:M17_pol_fulmap}.}
\label{fig:a3-closeup}
\end{figure*}

Figure \ref{fig:a3-closeup} defines the upper and lower regions used for the data analysis. We present in the figure different total intensity scales for better vision of the emission structure of the two regions.

\begin{figure*}[htbp!]
\centering
\includegraphics[width=1\textwidth]{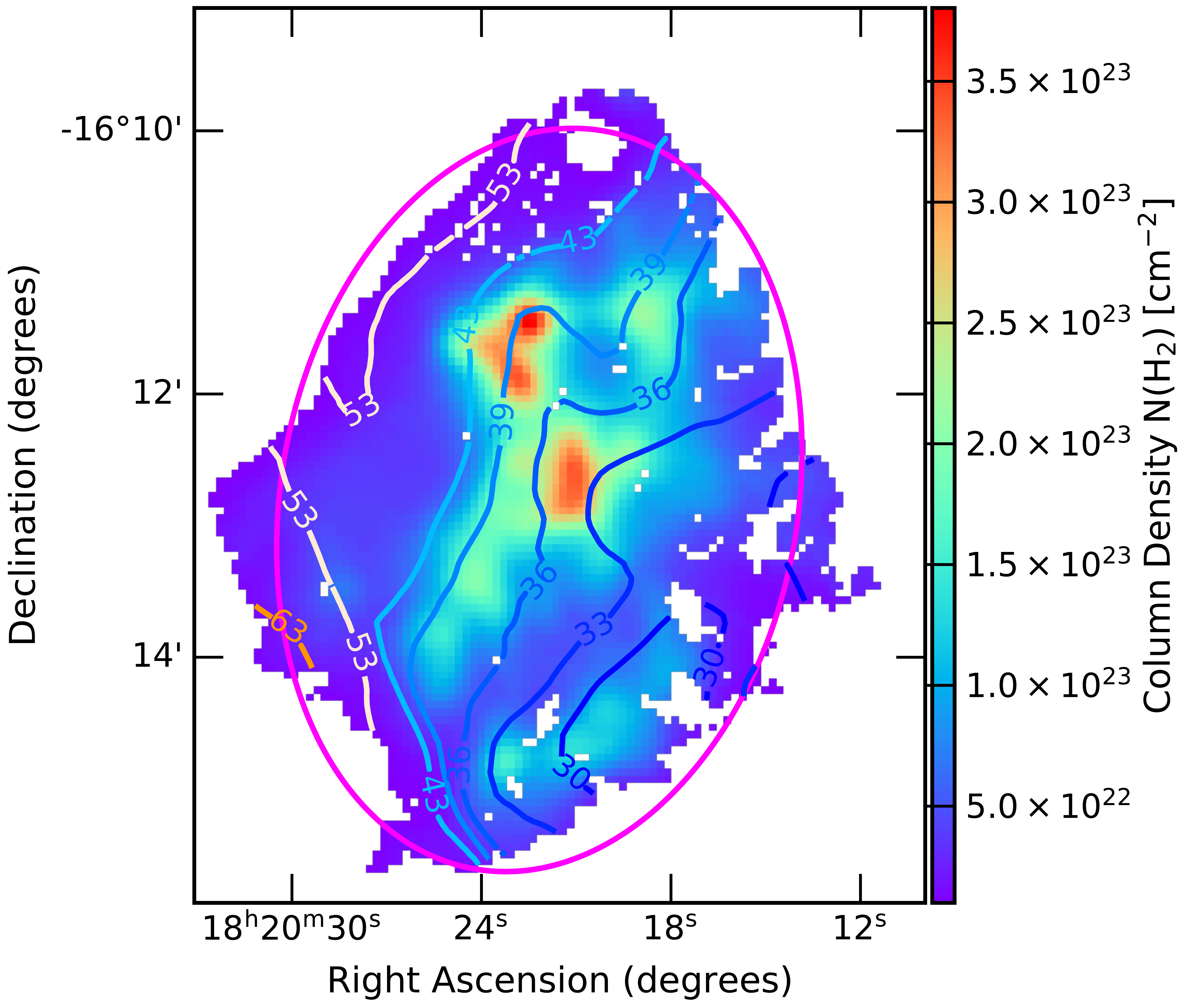}
\caption{The highest density region of the M17-S with $N(\rm {H_2}) > 10^{22}\;cm^{-2}$. The ellipse has a center at RA $\sim$ $\rm 18^h20^m22^s.17$ and DEC $\sim$ $\rm-16^\circ12\arcmin 48\arcsec.3$, major $\times$ minor axes of 171\arcsec $\times$ 118\arcsec, and a position angle of $350^\circ$. The contours are the dust temperature.}
\label{fig:a4-NH2}
\end{figure*}

Figure \ref{fig:a4-NH2} presents the highest density region of M17-S when applying a cut on $N(\rm {H_2}) > 10^{22}\;cm^{-2}$.

\end{document}